\documentclass{aa}
\usepackage{xcolor}
\usepackage{natbib}
\usepackage{graphicx}
\usepackage{stfloats}
\usepackage{placeins}
\usepackage[flushleft]{threeparttable}
\usepackage{tikz}

\usepackage{pdflscape}
\usepackage{rotating}
\usepackage{wrapfig}

\usepackage{txfonts}
\usepackage[breaklinks,colorlinks,urlcolor=blue,citecolor=blue,linkcolor=blue]{hyperref}
\begin{document}

   \title{Observing planetary gaps in the gas of debris disks}

   \author{C. Bergez-Casalou\inst{1} $\&$ Q. Kral\inst{1}}

   \institute{\inst{1}LESIA, Observatoire de Paris, Université PSL, CNRS, Sorbonne Université, Univ. Paris Diderot, Sorbonne Paris Cité, 5 place Jules
Janssen, 92195 Meudon, France, email: \texttt{camille.bergez@obspm.fr} \\}


 
 \abstract
  { Recent ALMA observations discovered consequent amounts (i.e., up to a few $10^{-1}\; \rm M_\oplus$) of CO gas in debris disks that were expected to be gas-free. This gas is in general estimated to be mostly composed of CO, C, and O (i.e., $\rm H_2$-poor), unlike the gas present in protoplanetary disks ($\rm H_2$-rich). At this stage, the majority of planet formation already occurred, and giant planets might be evolving in these disks. While planets have been directly observed in debris disks (e.g., $\beta$ Pictoris), their direct observations are challenging due to the weak luminosity of the planets. In this paper, with the help of hydrodynamical simulations (with \texttt{FARGO3D}) coupled with a radiative transfer code (\texttt{RADMC-3D}) and an observing tool (\texttt{CASA}), we show that planet-gas interactions can produce observable substructures in this late debris disk stage. While it is tricky to observe gaps in the CO emission of protoplanetary disks, the unique properties of the gaseous debris disks allow us to observe planetary gaps in the gas. Depending on the total mass of the gaseous debris disk, kinks can also be observed. We derive a simple criterion to estimate in which conditions gaps would be observable and apply it to the known gaseous debris disk surrounding HD138813. In our framework, we find that planets as small as $0.5 \; \rm M_J$ can produce observable gaps and investigate under which conditions (i.e., gas and planets characteristics) the substructure become observable with ALMA. The first observations of planet-gas interactions in debris disks can lead to a new way to indirectly detect exoplanets, reaching a population that could not be probed before, such as giant planets that are too cold to be detected by direct imaging.} 
  


   \authorrunning{C.Bergez-Casalou et al}
   \titlerunning{Observability of CO kinks in the gas of debris disks}

   \maketitle
%

\section{Introduction}

The gas-rich protoplanetary disks in which planets form have a limited lifetime due to different gas dispersal mechanisms (e.g., photo-evaporation and viscous accretion; see review by \citealt{Pascucci2023}). After the disk's dispersal, the resulting system is expected to be composed of the formed planets and can harbor dust and planetesimal disks, called debris disks. These disks are characterized by their dust emission and their old age ($\gtrsim$ 10 Myr). 

\begin{figure}[t]
        \centering   
        \includegraphics[scale=0.33]{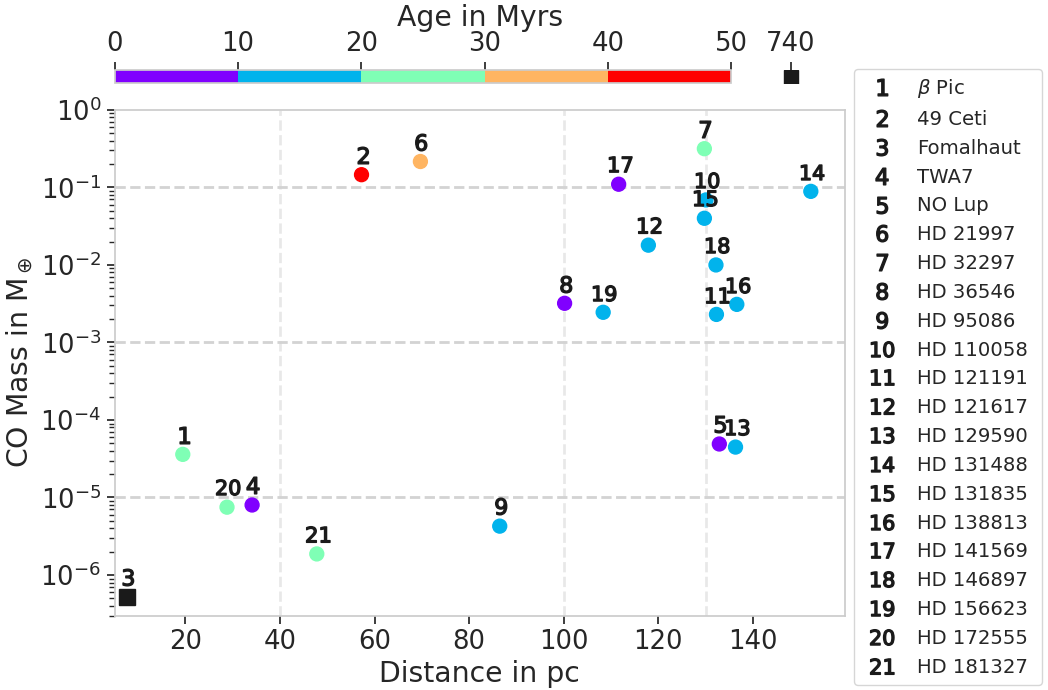}
        \caption{CO masses derived from CO observations as function of distance from the Sun. The color-scale represents the median estimated age of the system in million years. We note the particular case of Fomalhaut, which is particularly old compared to the other disks. The masses and distances investigated in this paper are marked by the horizontal and vertical dashed lines, respectively.}
        \label{mco_age}
\end{figure}

Consequential amounts of gas (mostly CO but also CI, CII, or OI) were recently observed around these old disks \citep[e.g.,][]{Riviere-Marichalar2012, Cataldi2014,Marino2016,Lieman-Sifry2016,Moor2017,Moor2019}. A summary of these gas observations can be found in Table \ref{tab:known_disks}. The amount of CO gas present in the most massive disks is estimated to be up to a few tenths of Earth masses (see Fig.\ref{mco_age}), which is comparable to the protoplanetary disk phase \citep{Zhang2021}. The survival of CO to photodissociation in these old disks can be explained thanks to a mechanism called shielding, where another component protects the molecule from external irradiation. In the case of CO, several shielding processes might be at play. For the youngest and most massive disks, they might still host consequent amounts of primordial gas \citep[called hybrid-disk gas, ][]{Kospal2013,Pericaud2017}, and $\rm H_2$ might be responsible for the shielding. However, it is not the favored scenario because, for the majority of the debris disks, the origin of the gas probably resides in the release of volatiles from collisional planetesimals \citep{Dent2014,Kral2017,Kral2019,Marino2020,Bonsor2023}. In this case, CO can be shielded by itself (self-shielding) and by an upper layer of neutral carbon; that is, CI \citep{Kral2019,Marino2020,Cataldi2023}.
The shielded gas can then viscously spread inward an outward from the planetesimal belt location \citep{Kral2016a, Kral2016b, Kral2019, Cui2024} and encounter already formed planets. This gas can be accreted by these planets and alter their atmospheres \citep{Kral2020Nat}. Similarly to protoplanetary disks, we expect the planet-gas interactions to produce different kinds of substructures. When a planet is embedded in a gaseous disk in differential rotation, it produces spiral arms originating from Lindblad resonances between the planet and the gas \citep[e.g.,][]{GoodmanandRafikov2001,ZhuZhang2022}. These spiral arms create both an overdensity in the disk and a deviation of the rotational velocity of the gas. These deviations from the usually Keplerian rotation of the gas produce observable structures in the gas emission close to the planet called kinks \citep[e.g.,][]{Teague2018,Pinte2018b}. 

When the planet becomes massive enough, the torque that it exerts on the gas can overcome the torque originating from the pressure and the viscosity of the gas, pushing the gas away from the planet's orbit \citep[e.g.,][]{Crida2006,Fung2014,Kanagawa2015,Bergez2020}. While this phenomenon has been intensively studied via hydrodynamical simulations, gaps are mostly observed in the dust emission of protoplanetary disks. As protoplanetary disks are gas-rich, the amount of gas remaining in gaps (even deep ones produced by massive giant planets) is sufficient for them to still be luminous, and therefore they are difficult to distinguish \citep[e.g., rare examples of gaps and cavities in][]{Keppler2019,Law2021}. However, as the gas around the planets is luminous, the kinks mentioned above become observable. On the other hand, gaps produce important gaseous pressure gradients able to trap dust at the edge of the gaps \citep[e.g.,][]{Pinilla2012,Zhang2018,Haugbolle2019}. Therefore, in protoplanetary disks, the gaseous components of gaps are hardly observable, but dusty gaps have been observed in the dust emission and some of them are consistent with a planetary origin. 


While observations already show that planets and gas can also co-exist in debris disks \citep[e.g., in the $\beta$ Pictoris system][]{Matra2017BetaPic,Lagrange2019}, models simulating the interactions between them only start to be developed \citep{Kral2020Nat}. In this study, we used hydrodynamical simulations to characterize planet-gas interactions such as gaps and kinks in the framework of debris disks. We also used a radiative transfer model and an observing tool in order to determine under which conditions these interactions are observable with ALMA. We focused on the CO emission as many debris disks are known to host this molecule in large quantities. This paper is structured as follows: the numerical setup of our different kind of simulations is presented in Sect. \ref{sec:numericalsetup}; in Sect. \ref{sec:results}, we show the synthetic images resulting from our parameter study; then, in Sect. \ref{sec:gapobserv}, we derive an observability criterion based on our results before discussing them in Sect. \ref{sec:discussion} and concluding in Sect. \ref{sec:conclusion}.

\section{Numerical setups}
\label{sec:numericalsetup}

In this study, we wanted to determine under which conditions substructures created by a giant planet embedded in a gas-rich debris disk can be observed by ALMA. To reach this goal, we proceeded in three steps: i) first, the gas distribution is determined by running hydrodynamical simulations with \texttt{FARGO3D}\footnote{\url{https://sites.google.com/view/pbllambay/fargo3d}} \citep{Benitez-Llambay2016}, taking different planet characteristics into account; ii) the emission of the resulting gas distribution is determined with the help of the radiative transfer code \texttt{RADMC-3D}\footnote{\url{https://www.ita.uni-heidelberg.de/~dullemond/software/radmc-3d/}} \citep{Dullemond2012}; iii) finally, realistic synthetic images are derived by adding noise and considering different ALMA configurations with the help of the observation simulation tool \texttt{CASA}\footnote{\url{https://casadocs.readthedocs.io/en/stable/}} \citep{CASA}. 

In this section, we present the different setups for each step, starting by presenting the different disk and planet configurations explored (Sect. \ref{subsec:config}). In Sects. \ref{subsec:hydro} and \ref{subsec:RT}, we present the numerical setups used for the hydrodynamical and radiative transfer simulations. The different ALMA configurations explored are developed in Sect. \ref{subsec:CASA}.

\subsection{Investigated configurations}
\label{subsec:config}

Gaseous debris disks can have a large range of different characteristics. In our study, we explore disks with different masses located at different distances, consistent with the already observed population. In Fig. \ref{mco_age}, we present some characteristics of all the known gas-bearing debris disks. A table with all the characteristics of each system is shown in Appendix \ref{appendix_knowndisks}. In order to match the characteristics of the observed disks, three different gas total masses are explored ($10^{-5} \rm \; M_\oplus$, $10^{-3} \; \rm M_\oplus$ and $10^{-1} \; \rm M_\oplus$, represented by the horizontal dashed lines in Fig.\ref{mco_age}), which are located at three different distances (40pc, 100pc, and 130pc; these are represented by the dashed vertical lines in Fig.\ref{mco_age}). 

Regarding the planet's characteristics, three different masses ($0.5 \; \rm M_J$, $ 1 \; \rm M_J$ and $5 \; \rm M_J$) located at three different semi-major axes (10 AU, 50 AU, and 100 AU) were explored. We focused on giant planets as they are the most likely to produce observable features while still being consistent with the currently known radial distribution of exoplanets \citep[e.g.,][]{Fulton2021,Zhu2022}.

\subsection{Hydrodynamical setup}
\label{subsec:hydro}

\begin{table}[t]            
\centering                          
\caption{Extent of different numerical grids depending on planet position.} 
\begin{tabular}{c c c c c}        
\hline             
  $r_p$       & $r_{min}$  & $r_{max}$  & $\Sigma_0(\rm 1 AU)$ & $m_{\rm disk}$   \\
  $\rm [AU]$  & $\rm [AU]$ & $\rm [AU]$ & $\rm [g/cm^2]$ & $[\rm M_\oplus]$  \\
\hline
10 AU   & 0.52      & 78        &  $5.4 \times 10^{-5}$ & $1.05 \times 10^{-3}$  \\
                 
50 AU   & 1.04      & 130       &  $3.1 \times 10^{-5}$ & $1.04 \times 10^{-3}$  \\
 
100 AU  & 29.64     & 200       &  $3.1 \times 10^{-5}$ & $1.36 \times 10^{-3}$  \\
\hline 
\end{tabular}
\tablefoot{The radial extent of the disks is chosen so that the planet is located far enough from the grid boundaries and the initial surface density profile is chosen so that the initial disk mass is $\sim 10^{-3} \; \rm M_\oplus$.}
\label{tab:numerical_grids}
\end{table}

In this work, we simulated the 2D ($r,\phi$) distribution of a gaseous disk surrounding a giant planet on a fixed circular orbit with the hydrodynamical code \texttt{FARGO3D} \citep{Benitez-Llambay2016}. The gas is represented by a single fluid, and the code follows its evolution by solving both the continuity and Navier-Stokes equations. The continuity equation reads
\begin{equation}
    \frac{\partial \rho}{\partial t} + \nabla . (\rho v) = 0,
\end{equation}

\noindent where $\rho$ is the volumic density of the fluid and $v$ its velocity. The Navier-Stokes equation solved in our setup is the following:

\begin{equation}
    \rho \left (  \frac{\partial v}{\partial t} + v.\nabla v \right) = - \nabla P + \nabla. T + F_{ext},
\end{equation}
\noindent where $P$ is the fluid pressure, $T$ is the stress tensor, and $F_{ext}$ represents external forces. The gaseous disk is then described by its surface density $\Sigma = \int \rho dz$ and its kinematic viscosity $\nu$. The equations are discretized over a 2D grid ranging from $r_{min}$ to $r_{max}$ with a radial resolution of $n_r = 704$ and from 0 to $2\pi$ with an azimuthal resolution of $n_\phi = 298$. The total size of the disk depends on the position of the planet (cf. Table \ref{tab:numerical_grids}). The disk's extent is chosen so that the planet is located far enough from the edges of the grid in order to keep nonreflective boundary conditions. The resolution is chosen so that at least five cells are present in the Hill radius $r_H$ of the planets (cf. Sect. \ref{subsec:CASA}).


The initial surface-density profile follows a power-law profile, with $\Sigma(r) = \Sigma_0 \times (r/\rm 1AU)^{-1}$ \citep[e.g.,][]{Kral2019}. $\Sigma_0$ is chosen so that the initial mass of the disk is roughly $10^{-3} \; \rm M_\oplus$. This value is then multiplied or divided by 100 to investigate the other two disk masses ($10^{-1} \rm \; M_\oplus$ and $10^{-5} \rm \; M_\oplus,$ respectively). Here, the disk is locally isothermal. The aspect ratio of the disk follows a power law where $h=H/r= 0.01 \times (r/\rm 1AU)^{0.25}$. With such profiles, the disk temperature also follows a power-law profile where $T = 171\rm K \times (r/\rm 1AU)^{-0.5}$, in line with what is expected from a second-generation gas in a debris disk \citep{Kral2019}. The disk's viscosity is described as in \cite{Shakura1973}: $\nu = \alpha c_s H = \alpha h^2 r^2 \Omega_K,$ where $\alpha$ is the turbulent viscosity parameter, $c_s = H \Omega_K$ is the isothermal sound speed, and $\Omega_K$ is the Keplerian frequency at a distance of $r$. Here, $\alpha$ is fixed to a value of $10^{-3}$, which is considered as a medium turbulent disk \citep[e.g.,][]{Cui2024}. As it can have a significant impact on the planet-gas interactions, we discuss this choice in Sect. \ref{sec:discuss_modelassump}.

For computational purposes, the code is based on dimensionless units. Masses are normalized to the mass of the central star, $M_0 = M_\odot$ and lengths to $r_0 = 5.2$ AU. The gravitational constant is $G=1$. The unit of time is therefore based on the orbital period at $r_0$ with $P = 2\pi t_0,$ where $t_0 = (r_0^3/(GM_{0}))^{1/2}$.

\begin{figure}[t]
        \centering   
        \includegraphics[scale=0.33]{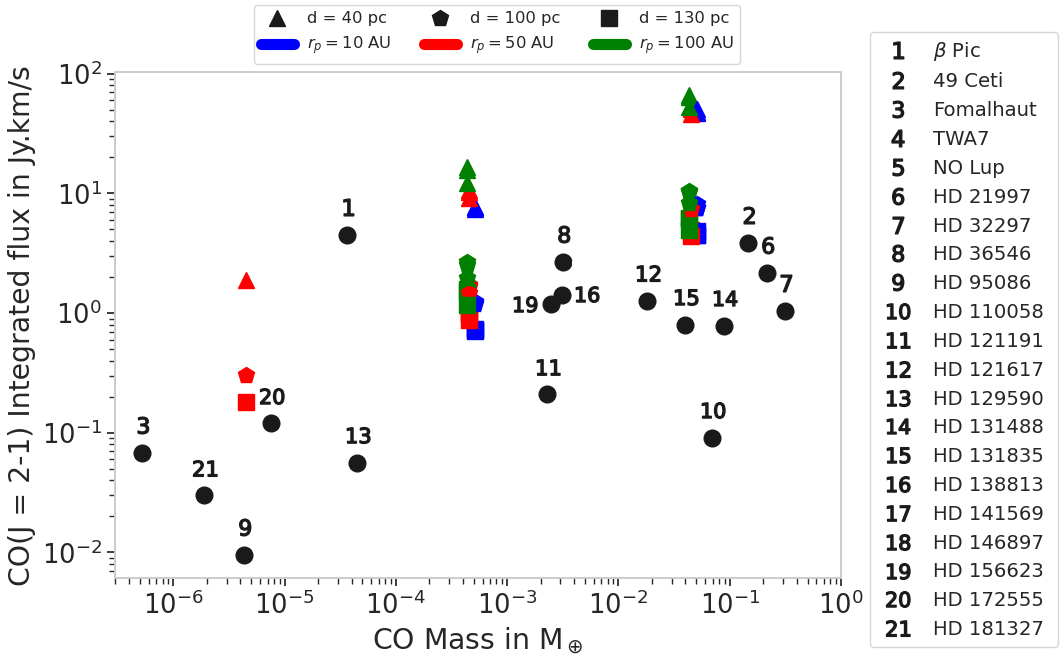}
        \caption{CO(J = 2-1) integrated fluxes as function of CO masses. The black dots represent the observations as in Fig. \ref{mco_age} and the colored symbols are derived from our simulations. The color represents the position of the planet and the symbols the distance of the disk. Our simulated disks are in line with the observed disks; the majority of them are located more than 100 pc away (cf. Fig. \ref{mco_age}). We only show the disks in one planetary configuration for the low-mass disk case ($m_{disk} = 10^{-5} \; M_\oplus$) as the disks are difficult to observe at high angular resolution in this case.}
        \label{mco_flux}
\end{figure}

The planets are slowly introduced in the disk with the following mass-taper function in order to prevent unrealistic shocks in the disk:
\begin{equation}
    m_{\rm taper} = \sin^2{(t/(4 n_{\rm orb}))}
    \label{masstapereq}
,\end{equation}
\noindent where $t$ is the time in dimensionless units. This function increases the planet's mass $m_p$ from zero to its final mass in $n_{orb}$ orbits at $r_0$. Here, we set $n_{orb}$ to 500 orbits for $m_p = 0.5 \; \rm M_J$, 3000 orbits for $m_p = 1.0 \; \rm M_J,$ and 5000 orbits for $m_p = 5.0 \; \rm M_J$, which is in line with gas accretion studies \citep[e.g.,][]{Hammer2017,Bergez2020}. In total, the disks are integrated for $10^4$ orbits at $r_0$ to reach a quasi-steady state and therefore obtain stable gaps.

\subsection{Radiative transfer setup}
\label{subsec:RT}

The gas emission is estimated using the radiative transfer code \texttt{RADMC-3D} \citep{Dullemond2012}. The 2D perturbed surface density and velocity distributions from \texttt{FARGO3D} are used to extrapolate the 3D distribution of the gas. Assuming vertical hydrostatic equilibrium \citep[e.g.,][]{ArmitageLN2022}, the volume density distribution follows:
\begin{equation}
    \rho(r,\phi,z) = \frac{\Sigma}{\sqrt{2\pi}\;H}\times \exp{\bigg(-\frac{z^2}{2H^2}\bigg)}
,\end{equation}
\noindent where $z = r\cos{\theta}$ ($\theta$ being the polar angle) resolved over $n_\theta = 16$ cells. The radial and azimuthal velocities are estimated to be constant with $\theta,$ and the vertical component of the velocity is null ($v_\theta$ = 0) as our simulation is isothermal and extrapolated from the 2D plane. The gas temperature is taken from the hydrodynamical setup. It is estimated to be entirely composed of CO with a molecular weight of $\mu = 28$. The disk is assumed to be in local thermodynamic equilibrium (LTE), given the range of disk masses we studied \citep[e.g.,][]{Matra2015,Kral2019}. Here, $n_{\rm photons} = 10^8$ photon packages are used.

The disk inclination is a crucial parameter for the observability of planet-gas interactions; face-on disks may seem ideal for the observations of gaps, whereas a low inclination of $\sim 30^\circ$ is needed in order to detect kinks \citep{Pinte2023}. In our fiducial setup, the disk inclination is $i = 30^\circ,$ and different inclinations from $i = 10^\circ$ to $i = 80^\circ$ are investigated in Sect. \ref{sec:diffinc}. 

The gas emission is calculated over a wide range of channel maps in order to properly scan the whole structure of the gaseous disk. 401 channels are simulated, from $dv = -20$ km/s to $dv = +20$ km/s centered on the CO(J = 2-1) emission line at 230.538 GHz. We chose our setup such that the resulting integrated fluxes are consistent with the observed disks. In Fig. \ref{mco_flux}, we show the total CO(J = 2-1) integrated flux of known gaseous debris disks as a function of the derived CO mass compared to our simulations. As expected, when located at 40pc, our disks are particularly luminous. However, the most massive disks observed are located at 100pc and further away, confirming that our simulations match the properties of the observed disks.

\begin{table}[t]            
\centering                          
\caption{Gap widths for planets located at various investigated semi-major axes with different masses.} 
\begin{tabular}{c c c c c}        
\hline             
        & 10 AU & 50 AU & 100 AU & $d_{disk}$   \\
\hline
        & 2.20 AU  & 11.0 AU & 22.0 AU & -      \\
$0.5 \; \rm M_J$  & 0.055"   & 0.275"  & 0.550"  & 40 pc  \\
        & 0.022"   & 0.110"  & 0.220"  & 100 pc \\
        & 0.017"   & 0.085"  & 0.169"  & 130 pc \\
\hline                  
        & 2.77 AU  & 13.9 AU & 27.7 AU & -      \\
$1 \; \rm M_J$  & 0.069"   & 0.347"  & 0.693"  & 40 pc  \\
        & 0.027"   & 0.139"  & 0.277"  & 100 pc \\
        & 0.021"   & 0.107"  & 0.213"  & 130 pc \\
 \hline                  
        & 4.74 AU  & 23.7 AU & 47.4 AU & -      \\
$5.0 \; \rm M_J$  & 0.119"   & 0.593"  & 1.186"  & 40 pc  \\
        & 0.047"   & 0.237"  & 0.474"  & 100 pc \\
        & 0.036"   & 0.182"  & 0.365"  & 130 pc \\
\hline
\end{tabular}
\tablefoot{ The gap widths are also given in arcseconds depending on the disk distance from Earth $d_{disk}$.}
\label{tab:gap_widths}
\end{table}

\subsection{Realistic image synthesis setup}
\label{subsec:CASA}

After being converted into Flexible Image Transport System (FITS) images compatible with \texttt{CASA}\footnote{\texttt{RADMC-3D} has a dedicated routine to produce FITS files compatible with \texttt{CASA.}} \citep{CASA}, we can add realistic noise to the gas emission in each channel map and convolve them with realistic beams corresponding to different ALMA configurations. 

The investigated ALMA spatial resolution is based on the width of the investigated gaps. From our hydrodynamical simulations (see Appendix \ref{appendix_hydro}), we see that the width of the gap corresponds to $w_{gap} = 4 r_H = 4 r_p (q/3)^{1/3}$ where $q$ is the planet-to-star-mass ratio, $r_p$ the planet's radius, and $r_H$ its Hill sphere radius \citep[e.g.,][]{Crida2006}. In Table \ref{tab:gap_widths}, we list the different gap widths in astronomical units and in arcseconds for the different disk distances investigated. We conclude that the optimal ALMA configuration is the C-6, corresponding to a beam of full width half maximum of $\sim 0.13"$ in Band 6. This configuration is combined with the C-3 configuration in order to correctly recover the total flux of the disk. In Sect. \ref{sec:gapobserv}, we discuss the observability of the gaps and the impact that different beam sizes can have on our results.

\begin{figure*}[t]
   \centering   
   \includegraphics[scale=0.35]{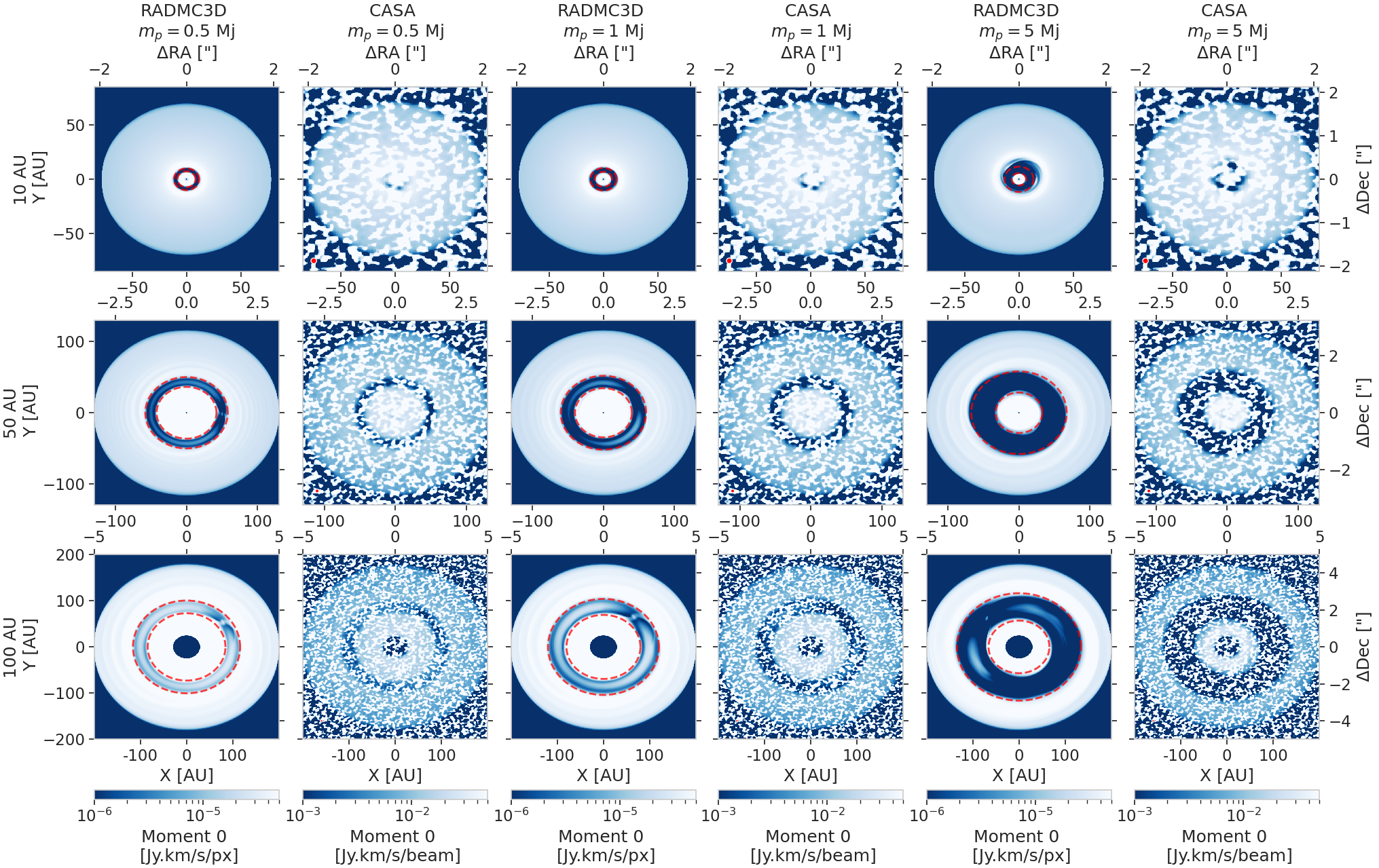}
   \caption{Moment-zero maps for three different planet masses (columns) located at three different semi-major axes (rows). We show the maps issued from \texttt{RADMC-3D} and \texttt{CASA}. The disk's mass is $\rm 10^{-3} \; M_\oplus$ of CO and located at 40 pc. The dashed red lines show the gap edges in the radiative transfer outputs. When the planet is located too close to the star (10 AU), the gap is hardly distinguishable given our ALMA resolution (beam of 0.13"). However, it is observed when the planet is > 50 AU and is more easily visible for $m_p > 0.5 \; \rm M_J$. }
   \label{moment0_diffplanetconf}
\end{figure*}

The interferometric visibilities are derived via the \texttt{simobserve} routine from the FITS images. Some thermal noise is added to each channel via the \texttt{CASA} routine \texttt{sm.setnoise(mode=simplenoise)}. The resulting noise corresponds to a noise of $\sim1.7$ mJy/beam/channel, taken each $dv = 0.1$ km/s. As in the \texttt{RADMC-3D} setup, the channels range from dv = $-20$ km/s to dv = $+20$ km/s centered on the CO(J=2-1) emission line, corresponding to a total bandwidth of 40 km/s (401 images in total). The resulting combined noise is therefore of 84 $\mu$Jy/beam. This sensitivity can be achieved in $\sim13$h on the source as estimated from the ALMA sensitivity calculator\footnote{\url{https://almascience.eso.org/proposing/sensitivity-calculator}}. The deconvolution is done via the \texttt{CASA tclean} routine via a Briggs weighting with robust value of 0.5. The complete \texttt{CASA} setup file is available upon request.

\section{Results}
\label{sec:results}

The resulting images in the different configurations are presented in the following sections. We present the emission from selected channel maps as well as the moment-zero maps in order to determine under which conditions the substructures are observable. The moment-zero maps are defined as follows:
\begin{align}
    \centering
    M_0 &= \sum_{i=1}^{N_{chan}} I_i  dv
,\end{align}
where $N_{chan}$ is the number of velocity channels, $I_i$ is the intensity per channel, and $d v$ is the velocity resolution.


\subsection{Influence of the planet configuration}
\label{sec:diffplanetmass}

As mentioned in Sect. \ref{subsec:config}, three different planet masses located at three different semi-major axes were explored in this study. These two characteristics govern the shape of the substructures that the planet will form in the gas disk. For example, a planetary gap is deeper and wider for a massive planet located far from its host star. This can be seen in Fig. \ref{moment0_diffplanetconf}, where we show a gallery of moment-zero images; here, the disks are $10^{-3}$ $\rm M_\oplus$ in mass, located at 40 pc, and host planets of different masses (increasing from left to right) and semi-major axes (increasing from top to bottom). The \texttt{RADMC-3D} outputs are shown next to the \texttt{CASA} images.

In the radiative transfer images, the planetary gaps (marked by the red dashed lines located at $r_p \pm 2r_H$) are always distinguishable. We can see that they are more or less emptied: some material still corotates with the planet, especially when the planet is located at 100 AU. This is expected as the gap depth depends on the local disk's scale height, which increases with radius in a flaring disk like ours (see Sect. \ref{sec:discuss_sigmagap}). One can note the most extreme case where the planet is $\rm m_p = 0.5 \; \rm M_J$ located at 100 AU: here, the gap is partially opened, with a lot of gas corotating with the planet. In this case, when the noise is added, the gap becomes barely distinguishable. However, at higher planetary masses, the gap is deeper and wider and therefore observable.

At the investigated resolution (0.13", marked by the small red dot in the bottom left corner of the \texttt{CASA} images), the gaps formed by the planets located at 10 AU become hard to distinguish due to the noise. The gaps are always circular, except in two cases: when the planet is massive ($\rm m_p = 5 \; \rm M_J$) or located closer to its host star (10 AU and 50 AU). This is due to the capacity of large planets to clear very deep gaps, reducing the damping of the planet eccentricity by the gas disk \citep[e.g.,][]{Papaloizou2001,Kley2006,Bitsch2010,Bitsch2013eccentricity}. 

\begin{figure}[t]
   \centering   
   \includegraphics[scale=0.45]{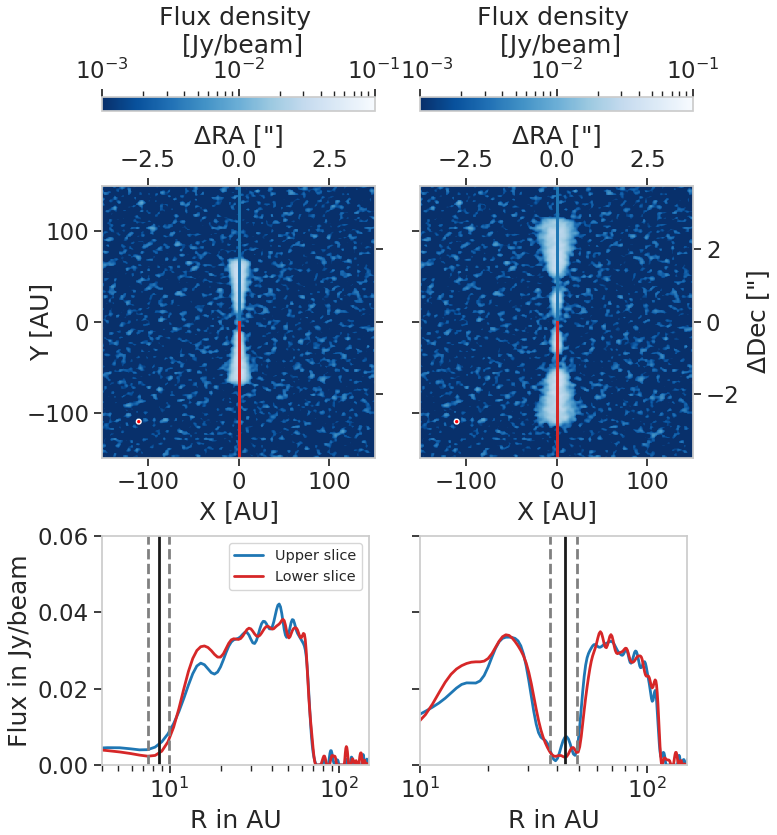}
   \caption{Channel maps at $dv = 0.0$ km/s (top panels) and two radial profiles of the gas emission along two radial axes (bottom panels). Here, the planet is $m_p = 1 \; \rm M_J,$ located, respectively, at 10 AU (left column) and 50 AU (right column). The upper and lower slices are the radial profiles taken as shown in the images. The vertical black line shows the planet location, and the dashed gray lines are located at $r_p \pm 2r_H$. Here, we clearly see that the gap is too close to the inner edge to be observed when the planet is located at 10 AU; however, the gap produced by the planet at 50 AU is wide and deep enough to be observed. }
   \label{profile_1em3}
\end{figure}

Another way to look at these gaps is to follow the radial profile of the emission along one of the radial axes of the disk. In Fig. \ref{profile_1em3}, we show two profiles taken from the channel maps at $dv = 0.0$ km/s. Two specific configurations are shown here: the planet is $1 \; \rm M_J$ and located at 10 AU (left column) and 50 AU (right column). All the other configurations can be seen in Appendix \ref{appendix_profiles}. As mentioned above, we clearly see here that the planetary gap produced by the planet at 10 AU is hardly distinguishable from the inner disk edge. However, when the planet is located further away, the gap is clearly marked. From such observations, one can measure the size of the gap and obtain a first estimate of the planet mass using the planet's Hill sphere:
\begin{equation}
    m_p = 3 M_* \left ( \frac{w_{gap}}{4 r_p} \right )^3
    \label{eq_gap_width}
,\end{equation}

\noindent where $w_{gap}$ is the gap width, $M_*$ the mass of the central star, and $r_p$ the planet position being in the center of the gap. We note that one needs to take into account the inclination of the disk in order to properly determine the gap width. In Fig. \ref{profile_1em3}, the position of the planet and the gap width are projected onto the sky, taking the disk's inclination into account ($r_{p,proj} = r_p \cos{(i)}$). We discuss the strength of the link between the gap width and the planet's Hill sphere in Appendix \ref{appendix_hydro}.

These images already show that, in the low-surface-density context of debris disks, planetary gaps are observable. Even if they are relatively as deep as in the protoplanetary-disk phase (see Sects. \ref{sec:diffdiskmass} and \ref{sec:discuss_sigmagap}), the absolute surface density is way lower. In the gap, the amount of gas is too low to be detected, its emission remaining below the sensitivity threshold of ALMA, while the other parts of the gas disk are bright enough to be detected. Therefore, the important parameter in determining whether gaps are observable or not is the absolute amount of gas present in the gap. In the next section, we test this assumption by increasing the total mass of the disk.

\begin{figure*}[t]
   \centering   
   \includegraphics[scale=0.39]{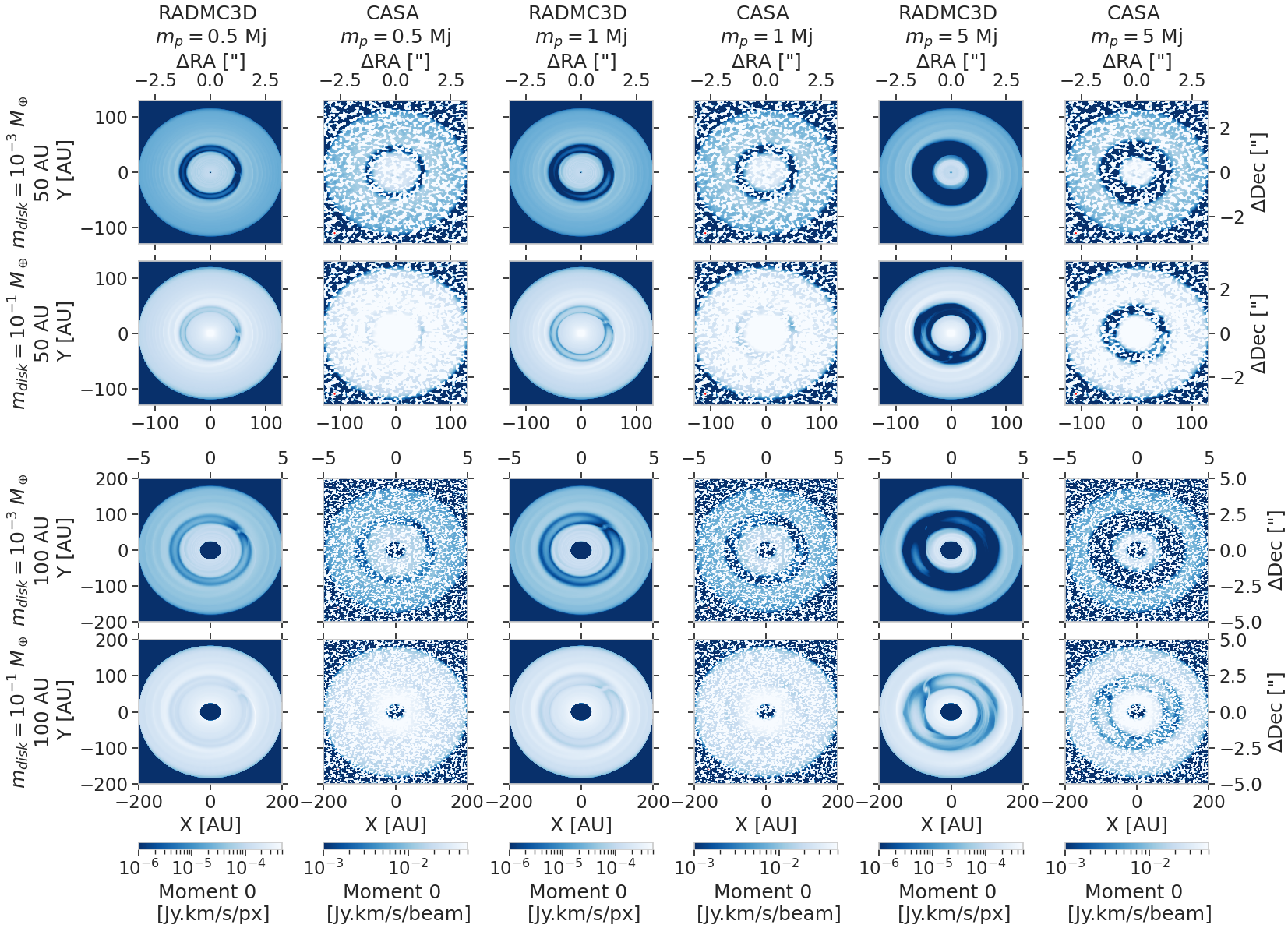}
   \caption{Moment-zero maps for two different disk masses hosting planets located at two different semi-major axes (rows) for three different planet masses (columns). We show the maps issued from \texttt{RADMC-3D} and \texttt{CASA}. The disks are located at 40 pc. The disk's mass is $\rm 10^{-3} \; M_\oplus$ of CO in the first and third rows and $\rm 10^{-1} \; M_\oplus$ in the second and last rows. As expected, the more massive disks are more luminous. The gas present in the gap can become luminous enough to prevent a clear detection of the gap, as in the lower disk-mass case.}
   \label{moment0_diffdiskmass}
\end{figure*}

\begin{figure*}[t]
   \centering   
   \includegraphics[scale=0.39]{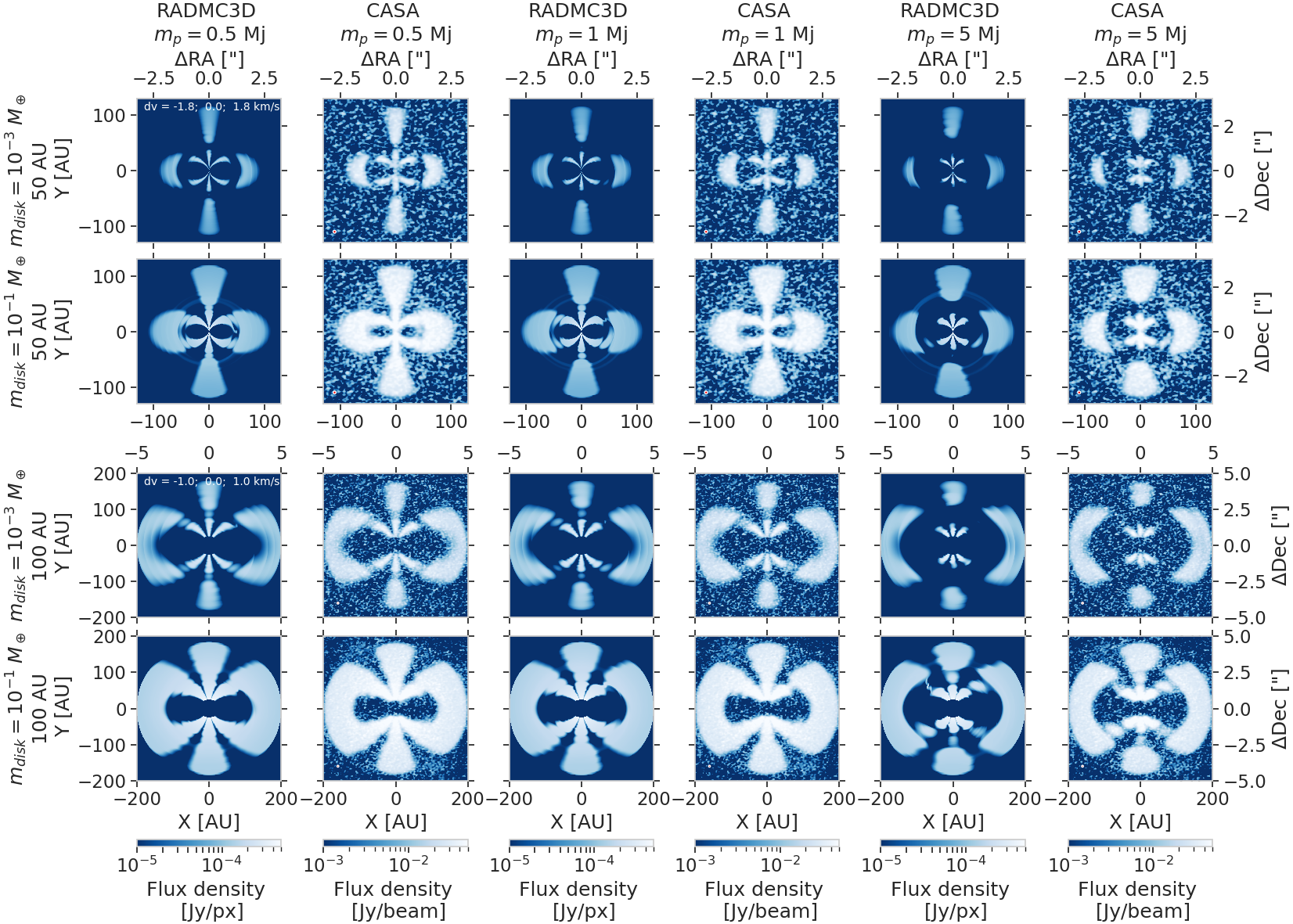}
   \caption{Combined channel maps at dv = -1.8; 0; 1.8 km/s (two first rows) and dv = -1.0; 0; 1.0 km/s (two last rows) for two different disk masses hosting planets located at two different semi-major axes (rows). The planet mass increases in the columns from left to right. We show the maps issued from \texttt{RADMC-3D} and \texttt{CASA}. The disks are located at 40 pc. Gaps are observed in low-mass disks, and kinks (marked by a red circle) are observed in high-mass disks.}
   \label{channelmaps_diffdiskmass}
\end{figure*}


\subsection{Influence of the disk mass}
\label{sec:diffdiskmass}

Gaseous debris disks can host gas with a wide variety of masses, ranging over several orders of magnitude (see Fig. \ref{mco_age}). In this section, we investigate the impact of the total disk mass on the observability of the different substructures. As mentioned in Sect. \ref{subsec:config}, three different disk masses were investigated in our study: $10^{-5} \rm \; M_\oplus$, $10^{-3} \; \rm M_\oplus,$ and $10^{-1} \; \rm M_\oplus$. However, due to the very faint emission of the $10^{-5} \rm \; M_\oplus$ case, which is difficult to correctly image with a realistic setup, we focused on the other two masses. 

In Fig. \ref{moment0_diffdiskmass}, we show the moment-zero images as in Fig. \ref{moment0_diffplanetconf} for the different planet masses investigated that are located two distances from the star. The first two rows show the planets located at 50 AU in the low-mass disk ($10^{-3} \rm \; M_\oplus$, first row) and in the most massive disk ($10^{-1} \rm \; M_\oplus$, second row). The same images are shown in the last two rows for the planets located at 100 AU. As mentioned previously and discussed in Sect. \ref{sec:discuss_sigmagap}, the relative gap depths are the same whether the planet is located in the $10^{-3} \rm \; M_\oplus$ or $10^{-1} \rm \; M_\oplus$ disk as it is independent of the surface-density profile \citep{Crida2006,Fung2014,Kanagawa2015}. The only difference is the absolute amount of gas present in the gap.

Some gaps that are easily distinguishable in the low-mass disk become too luminous to be distinguished from the emission outside the gaps, as in the protoplanetary disk case. This can be seen in the low-mass planet cases at both distances; while the disk hosting the $m_p = 0.5 \; \rm M_J$ planet was observable at 50 AU, it becomes hard to distinguish in the high-disk-mass case. This can also be noted for the cases of $m_p = 1 \; \rm M_J$ at 50 and 100 AU, though it is less distinct. Therefore, the absolute amount of gas remaining in the disk is the main characteristic governing whether one can observe planetary gaps in gaseous disks. We develop this aspect in Sect. \ref{sec:gapobserv}.

When the emission inside the gap becomes significant, the configuration tends to be more similar to that of a protoplanetary disk (see also Appendix \ref{appendix_profiles}). However, while the gap size itself is hard to determine, other features become observable, such as kinks \citep{Pinte2018b,Teague2018}. These features originate from the impact that planets have on the normally Keplerian rotation of the disk; that is, the embedded planet launches spiral waves that can either accelerate or decelerate the Keplerian rotation of the unperturbed disk \citep[for a review, see][]{Pinte2023}.

These deviations can be seen in the emission of the individual channel maps (Fig. \ref{channelmaps_diffdiskmass}). In a non-perturbed disk, the lobe-shape produced by the Doppler-shifted emission from the disk is perfectly symmetrical. As the deviation from the planetary spiral wave is stronger near the planet location, it produces an asymmetry in the emission. 

In this study, we show, for the first time, that these kinks are also observable in the most massive debris disks; in Fig. \ref{channelmaps_diffdiskmass}, the combined emission of three different channel maps is shown for the same configurations as in Fig. \ref{moment0_diffdiskmass}. Here, we can see that when there is too much gas in the corotation region of the planet, the emission from the different channels is in a continuous-lobe shape, meaning that the gap is not observed. We now give an example. In the case where $m_p = 1 \; \rm M_J,$ located at 100 AU or 50 AU, the gap is distinguishable from the moment-zero map (Fig. \ref{moment0_diffdiskmass}) and can be noticed on the different channel maps in Fig. \ref{channelmaps_diffdiskmass} in the low-mass disk cases. However, when the mass of the disk is increased, the gap is barely distinguishable. On the other hand, a kink can be seen in the channel dv = 1 km/s or 1.8 km/s, probing the emission of the gas located closest to the planet. The same process happens in the $m_p = 5 \; \rm M_J$ case located at 100 AU, and here the spiral wave is so strong that it produces a kink visible in different channel maps (i.e., further away from the planet).

On the other hand, seeing the gap in the low-mass disk does not necessarily mean that a kink is visible in the high-disk-mass case, as can be seen when the planet is $m_p = 0.5 \; \rm M_J$ located at 50 AU. This is expected as, for the same planet mass and location, the gap width is larger than the width of the spiral arms producing the deviation from the rotational speed, requiring a higher spatial resolution to observe the kink compared to the gap. Moreover, we have one case where no gap is seen in the low-mass disk and no kink is seen in the high-mass disk ($m_p = 0.5 \; \rm M_J$ located at 100 AU) due to the small impact the planet has on the disk (i.e., no deep gap and small spiral arms). In this particular case, a small kink starts to be distinguishable in the low-mass disk.

From this parameter study, we conclude that the planets located in gaseous debris disks can produce two kinds of observable substructures: gaps or kinks. The observability of each structure depends on the emission of the absolute amount of gas in the corotation region of the planet compared to the ALMA configuration thresholds. In order to see gaps, the planet must create a deep enough gap in a low-mass disk, while kinks are observable only for the high-mass planets located in massive disks. We outline a criterion in Sect. \ref{sec:gapobserv} that allows us to estimate when gaps are observable.

\subsection{Influence of the inclination}
\label{sec:diffinc}

\begin{figure*}[t]
   \centering   
   \includegraphics[scale=0.47]{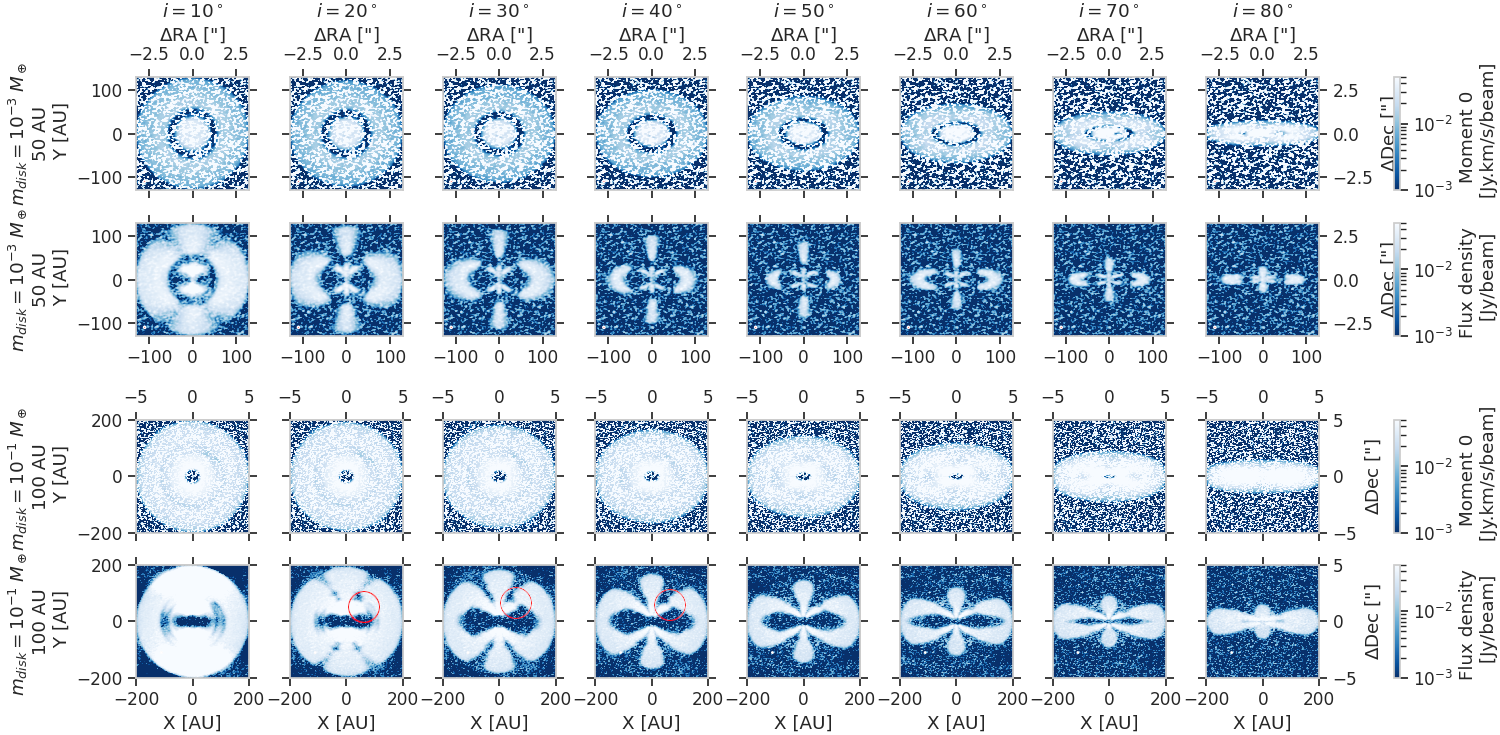}
   \caption{Moment-zero and combined channel maps at different $dv$ for two different configurations. Here, the selected channels are the ones where the gas emission is located closest to the planet in order to clearly see its impact on the gas. In the first top rows, a Jupiter-mass planet located at 50 AU is hosted in a $10^{-3} M_\oplus$ disk, producing a gap; while in the two last rows, a Jupiter mass planet located at 100 AU is introduced in a more massive disk ($10^{-1} M_\oplus$), producing a kink. The disks are located at 40 pc and are inclined with increasing inclination from left ($i=10^\circ$) to right ($i=80^\circ$). The visibility of the kink (traced by the red circle) is highly influenced by the disk's inclination, unlike the gaps.}
   \label{maps_diffdiskinc}
\end{figure*}

The observed debris disks are known to have a wide range of inclinations, ranging from $\sim 10^\circ$ to almost edge-on disks (see Table \ref{tab:known_disks}). However, it is known that the inclination of the disks can significantly impact the observability of gaseous substructures \citep[e.g.,][]{Pinte2018a,Barraza2024}. In this section, we investigate eight different inclinations and their impact on both gaps and kinks produced by our planets. In Fig. \ref{maps_diffdiskinc}, we show the simulated \texttt{CASA} images in two specific configurations (a $1 \; \rm M_J$ planet at 50 AU in a low-mass disk in the top rows, producing an observable gap; and a $1 \; \rm M_J$ planet at 100 AU in a high-mass disk in the bottom row, producing a kink) in disks with various inclinations. In order to match the characteristics of the observed disks, the inclination ranges from $10^\circ$ to $80^\circ$. 

Focusing first on the low-mass-disk case, the gap induced by the Jupiter-mass planet is more easily observable in disks at low inclination. This originates from the fact that when the disk is inclined, part of the disk that is closer to the observer hides the gap. However, the emission along the semi-major axis of the inclined disk is only slightly affected by this effect \citep{Bergez2022}. Therefore, even if the disk is highly inclined, the gap is still observable, especially along the semi-major axis. The impact of the inclination on the gap observability also depends on the gap depth: a narrow gap will be easily hidden along the semi-minor axis of the disk, while a wide gap will still be distinguishable along all axes. 

The inclination has a more important impact on the observability of the kink. Indeed, the shape of the gaseous emission in individual channel maps highly depends on the projected velocity of the gas \citep[e.g.,][]{Rosenfled2013,Pinte2018a}. The Keplerian rotation of disks at intermediate inclination (here estimated to be $20^\circ < i < 45^\circ$) allows us to spatially separate different emitting regions. As kinks are defined as deviation from this Keplerian rotation, we need to be able to spatially resolve the distortion of the usually symmetrical lobe-shaped emission. It is clear from Fig. \ref{maps_diffdiskinc} that the kink produced by the $1 \; \rm M_J$ planet is stronger for $ i = 20^\circ$ and $ i = 30^\circ$. However, one can still distinguish it at $ i = 10^\circ$ and $ i = 40^\circ$. Due to the noise and the shape of the Keplerian emission, the deviation is not observable for higher inclinations. Therefore, we conclude here that the disk inclination mainly impacts the observability of kinks, as expected from studies of protoplanetary disks. On the other hand, the observability of the gap is only slightly impacted: it also depends on the width of the gap, with wider gaps being less impacted than narrow ones. This is encouraging regarding a future observational search for gaps in debris disks as there are no strong constraints on the inclination distribution. 

\section{Estimating the observability of the gap}
\label{sec:gapobserv}

In the previous section, we show that planetary gaps can become observable in the gas emission of debris disks. In this section, we derive an analytical criterion allowing us to estimate under which circumstances a gap is visible; this depends on the properties of the system and the observational configuration. 

As presented in Sect. \ref{sec:diffplanetmass}, the first step is to determine whether the gap is resolved. So far, we present images of disks located at 40 pc, corresponding to the closest debris disks observed (see Fig. \ref{mco_age}). We also investigated the observability of the substructures in disks located at 100pc and 130pc (see Appendix \ref{appendix_distances}), where it becomes clear that the gap needs to be wide enough to be covered by more than one beam size (see also Sect. \ref{sec:HD138813}). Then, one can use the approach detailed in Sect. \ref{subsec:CASA} based on the Hill sphere of the planet to determine the size of the gap.

Once we know that the gap is resolved, we need to estimate the emission of the gas inside the gap and compare it to the instrument's sensitivity. In order to do so, we estimate the flux $\rm F_{u,l}$ emitted by a parcel of mass $\rm M_g$ by
\begin{equation}
    F_{u,l} = \frac{h\nu_{u,l} A_{u,l} x_u M_g}{4\pi d^2 m_{mol}}
    \label{eq_flux}
,\end{equation}
\noindent where $h$ is the Planck constant, $\nu_{u,l}$ the frequency of the transition from the upper level $u$ to the lower level $l$, $d$ the distance, and $m_{mol}$ the mass of the studied atom or molecule. $x_u$ is the fraction of molecules that are in the upper level $u$ and can be expressed as a function of the partition function $Z$:
\begin{equation}
\begin{split}
        & Z = \sum_{i} g_i e^{-E_i/k_BT_g} , \\
        & x_u = \frac{g_u}{Z}e^{-E_u/k_BT_g}
\end{split}
,\end{equation}

\noindent where $g_i$ is the degeneracy of level $i$, $k_B$ the Boltzmann constant, and $T_g$ the gas temperature \citep{Matra2015}. Here, the central frequency is $\nu$ = 230 GHz for the CO(J = 2-1) line and $m_{mol} = m_{CO} = 28 \; m_p$, where $m_p$ is the proton mass.



It is possible to estimate the integrated flux inside the gap by deriving the gas mass in the gap from our hydrodynamical simulations. As the gap is spatially resolved, we consider the mass contained in one beam of area $S_{beam}$, resulting in the mass $M_g = S_{beam} \Sigma_{gap}$. The flux in one beam located inside the gap is therefore

\begin{equation}
    F_{gap} = \frac{h\nu_{u,l} A_{u,l} x_u S_{beam}\Sigma_{gap}}{4\pi d^2 m_{mol}}
    \label{eq_fluxcriterion}
.\end{equation}

\begin{figure}[t]
   \centering   
   \includegraphics[scale=0.63]{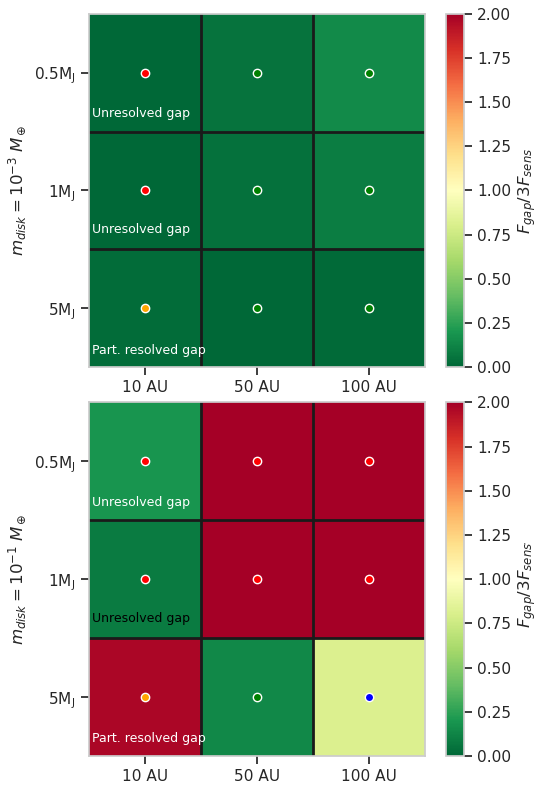}
   \caption{Value of our observability criterion $F_{gap}/(3 F_{sens})$ (color map) in different configurations explored. Here, the disks are located at 40 pc. Both disk masses are shown in the top (low-mass disk) and bottom (high-mass disk) squares. Each panel of each square represents a planetary configuration, with the planet mass increasing from top to bottom and its semi-major axis increasing from left to right. The dots represent whether we actually observe a gap (green if observed, red if not) in the simulated images presented in the previous sections. If our criterion works, then the color of the dot matches the color map. Here, we see that our criterion matches our simulated images quite well, with a few exceptions (see the text). Some gaps are either eccentric (orange dots) or observed but luminous (blue dot), which can complicate our analysis (see text). }
   \label{observability_map}
\end{figure}

This value is to be compared to the ALMA sensitivity $F_{sens}$, based on the noise in the moment-zero maps (see Sect. \ref{subsec:CASA}). In order to have a 3 $\sigma$ detection, we estimate that the flux in the gap has to be higher than $3 F_{sens}$. Consequently, in order for the gap to be detectable, the flux from the gap has to be smaller than $3 F_{sens}$. Here, $3F_{sens}$ is measured to be $3F_{sens} \simeq 12.9$ mJy.km/s for our 401 channels ranging from -20 km/s to 20 km/s.

In Fig. \ref{observability_map}, we show the ratios $F_{gap}/(3 F_{sens})$ where $F_{gap}$ is the flux determined from Eq. \ref{eq_fluxcriterion} in different configurations: each square containing nine panels represents a disk mass ($\rm m_{disk} = 10^{-3} \; M_\oplus$ in the top square and $\rm m_{disk} = 10^{-1} \; M_\oplus$ in the bottom square), and each panel represents a planet configuration, with the mass increasing from top to bottom and semi-major axis increasing from left to right. The gap is estimated to be observable when the value of the $F_{gap}/(3 F_{sens})$ ratio is smaller than 1 (shown in green). Overlaid are dots representing what we find from our images. A green (resp. red) dot means that we estimate that the gap is (resp. not) distinguishable from the results shown in Sect. \ref{sec:results}. If our criterion works efficiently, the color of the background matches the color of the dots.

We see from Fig. \ref{observability_map} that our criterion works quite well; the gaps are observable mostly in the low-mass disks and are hard to observe in the high-mass case. We now focus on the peculiar cases; first, the gaps located at 10 AU are not spatially resolved, meaning that the first step of our criterion is not met here, given our ALMA resolution. Therefore, for those cases, the calculation of the flux ratio is not relevant. In the particular case where the planet is $5 \rm M_J$ located at 10 AU, the gap is at the limit of the angular resolution (see Table \ref{tab:gap_widths}). Moreover, in this case, the gap starts to become eccentric. Both of these effects make the estimation of the gap observability sensitive, as the gap becomes partially resolved and can therefore be partially observed.

Another interesting configuration is when the planet is located at 100 AU in the high-disk-mass case (marked with the blue dot). In Fig. \ref{moment0_diffdiskmass} we see that the gap is distinguishable thanks to its large width, even though there is still a significant amount of gas inside the gap (Fig. \ref{moment0_diffdiskmass}). Therefore, the ratio $F_{gap}/(3 F_{sens})$ is close to one (yellowish color). Here, what makes the gap observable is the fact that the amount of gas inside the gap is relatively less luminous compared to the outer part of the disk. Therefore, even if the presence of the gap might be ambiguous from the moment-zero image, it should be seen in the radial profile, meaning that our criterion is correct here too. 

We conclude that our criterion (estimated from the ability to resolve the gap width and Eq. \ref{eq_fluxcriterion}) is an efficient way to determine whether gaseous gaps can be observed in the gas emission of disks. We note that this criterion does not depend on the nature of the disk itself; the only requirements are that the gas is luminous enough outside of the gap to be detected, the gap can be spatially resolved, and $F_{gap}<(3F_{sens})$. Therefore, this criterion can also be applied to low-mass protoplanetary disks.

\section{Discussion}
\label{sec:discussion}

\subsection{Application to a known debris disk: HD138813}
\label{sec:HD138813}

In the previous sections, we show that planetary gaps are observable in disks representative of the observed debris-disk population. Here, we apply our approach to an ideal candidate to investigate if we can observationally validate this study. In order to do so, we require three main critera to be fulfilled: i) the disk must be luminous enough to be detected outside the gap, while the gap must be deep enough to push the gas emission below ALMA's sensitivity; ii) the disk cannot be too inclined to prevent the gas from hiding the gap; and iii) the width of the gap must be spatially resolved. From our parameter study, we estimate that the CO gas mass must therefore be $10^{-4} \; \rm M_\oplus < m_{disk} < 10^{-2} \; \rm M_\oplus$ and that the inclination must be lower than $60^\circ$.

\begin{figure}[t]
   \centering   
   \includegraphics[scale=0.34]{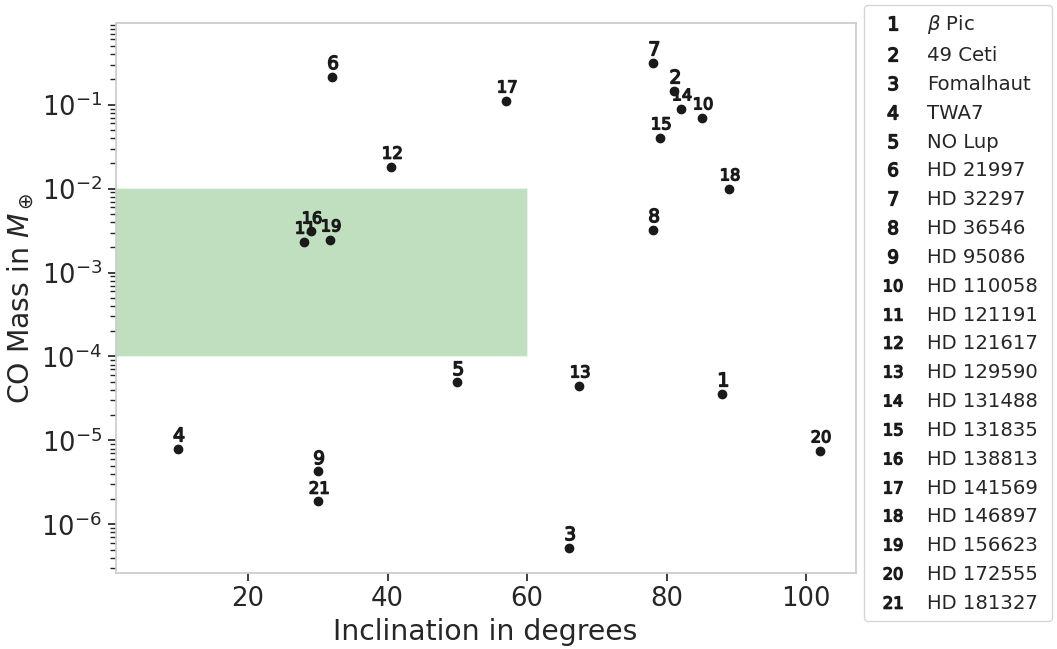}
   \caption{Debris disks with CO detections. The disk's CO mass and inclination are used to determine the ideal candidates (green area); from our simulations, if the disk is too massive (i.e., luminous), the emission in the gap is too big, similar to the protoplanetary disk regime. On the other hand, if the disk is too light, it is not luminous enough to be observed at a high resolution. Moreover, a too inclined disk prevents a clear detection of the gap.}
   \label{co_mass_inc}
\end{figure}

In Fig. \ref{co_mass_inc}, we show the mass/inclination distribution of the debris disks known to host CO gas; the ideal candidates regarding the first two criteria are shown in the green area (with $10^{-4} \; \rm M_\oplus < m_{disk} < 10^{-2} \; \rm M_\oplus$ and $i<60^\circ$). The ideal candidates are HD121191, HD138813, and HD156623. The last important step is to estimate the width of planetary gaps in order to determine if they can easily be spatially resolved with ALMA or not.

If a planet is indeed embedded in these disks, it would be more likely located at the edge of the dust belt: during the protoplanetary-disk phase, the dust could have been accumulated at the edge of the planet gap, where it could grow to planetesimal sizes \citep[e.g.,][]{Morbidelli2020}. After the dispersal of the primordial gas, the collisions in the planetesimal belt would release dust and gas to the extent that we observe nowadays. Assuming the planet location $r_p$ at the edge of the dust disk, the width of the gap $w_g$ can then be estimated from its Hill radius $r_H$, as shown in Sect. \ref{subsec:CASA}. In the case of HD121191 \citep[located at 132.3pc with its inner dust edge at 21.5 AU][]{Kral2020}, a Jupiter mass planet would produce a gap of 0.045" ($\sim$ 6 AU), which would require a very high angular resolution to resolve it. A more massive planet --for example, a 5 Jupiter-mass planet-- would produce a wider gap of course (i.e., $\sim$10 AU corresponding to 0.077"), but it can be expensive to catch with ALMA (i.e., with the highest angular resolutions, the integration time to collect all the fluxes can be very high). For HD156623 \citep[located at 108.3 pc with its inner dust edge at 26 AU][]{Lieman-Sifry2016}, the same problem occurs: a Jupiter (resp. 5 Jupiter) mass planet would produce an $\sim$7 AU gap, corresponding to 0.06" (resp. $\sim$12 AU or 0.11"). However, in the HD138813 disk, located at 136.6 pc, the inner dust edge is estimated to be located farther inside the disk \citep[$\sim$ 70 AU][]{Hales2019}. This results in wider gaps; for a Jupiter-mass planet, the gap is $\sim$20 AU wide, corresponding to 0.14" and a 5 Jupiter-mass planet would produce an $\sim$33 AU gap (0.24"). This becomes easily resolvable with ALMA, for example in the C-7 configuration (0.092"). Unfortunately, archived data of this disk do not provide a high enough resolution to test this hypothesis directly. Follow-up observations are required. 

In the case of HD156623 and HD121191, the gap width is small as we made the assumption that the planet would be located at the inner edge of the debris disk. However, when observed at high angular resolution, some debris disks are known to host rings and gaps in their dust distribution \citep{Ricci2015,Marino2018}. HD156623 and HD121191 might also host such gaps in their dust distribution; if future high-angular-resolution observations of the dust emission of these disks show gaps, then it would be natural to place planets in them and look for the gaseous counterpart of the gap (see Sect. \ref{sec:discuss_gapdust}). In this case, the planets can be located further away in the disk and produce wider gaps, making them easier to detect.

In order to derive synthetic images of HD138813 at a higher angular resolution, we followed these steps: first, hydrodynamical simulations based on the observed disk's characteristics (see Table \ref{tab:known_disks}) are run in three different configurations (with the disk only, hosting a Jupiter or a 5 Jupiter-mass planet). Using Eq. \ref{eq_fluxcriterion}, we can estimate whether the gap should be observable or not: from the hydrodynamical simulation, we find that $\Sigma_{gap} = 1.49 \times 10^{-8}$ $\rm g/cm^2,$ resulting in a flux in the gap of $F_{gap} = 5.41 \times 10^{-4}$ Jy.km/s. Using the same setup as in this study except in the C-7 configuration, the resulting integrated noise is approximately $\sim 4 \times 10 ^{-3}$ Jy.km/s, meaning that $F_{gap} \ll 3 F_{sens}$. Therefore, the gap should be observable.

\begin{figure}[t]
   \centering   
   \includegraphics[scale=0.37]{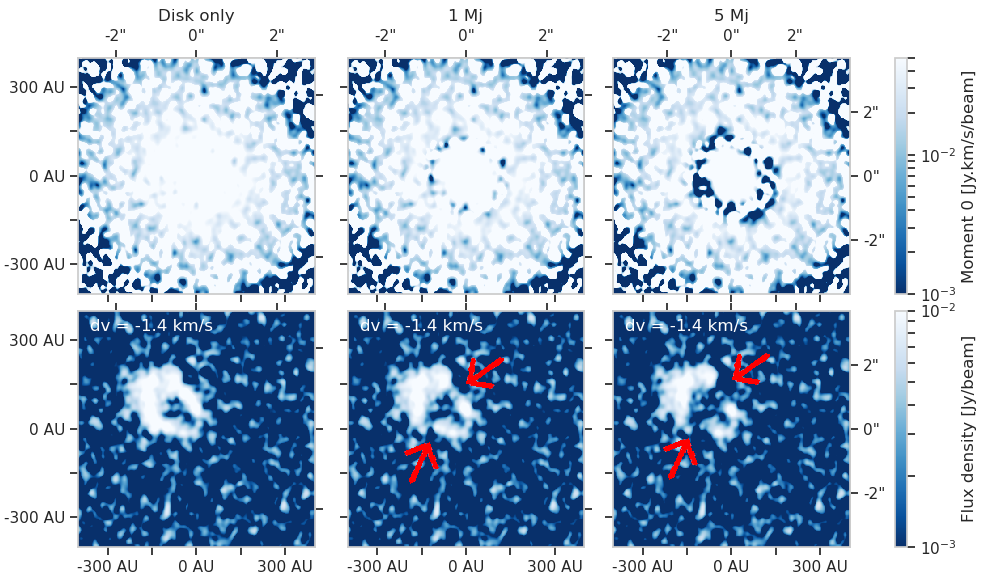}
   \caption{Simulated images of CO(J = 2-1) emission of HD138813 in three different configurations. In the first column, the disk does not host any planet and is to be compared to the two other columns where a Jupiter-mass planet (center) and a five-Jupiter-mass planet (right) have been introduced in the disk at 70 AU. Moment-zero maps are shown in the top row, and single-channel maps at $dv = -1.4$ km/s are shown in the bottom row. In the smooth-disk case, the lobe shape emission is continuous, whereas the gap produced by the planets cuts the emission, making the gap observable.}
   \label{image_HD138813}
\end{figure}

In Fig. \ref{image_HD138813}, we show the simulated images in the case without a planet and with a Jupiter or 5 Jupiter-mass planet. It is clear from the moment-zero maps that the gap is distinguishable when hosting planets, meaning that Jupiter-mass planets create deep enough gaps in this disk. We also note here that it might be easier to derive the size of the gap from the individual channel maps than from the moment-zero image. Here, the total integration time is $\sim$ 7h on source.

In this section, we show that the observation of gaseous gaps in debris disk is a realistic approach that just needs to be tested observationally. Future high-angular observations of gaseous debris disks will therefore help us understand the process of gap formation, as gaseous planetary gaps are barely observable in protoplanetary disks.

\subsection{Influence of the disk's characteristics}
\label{sec:discuss_modelassump}

For the first time, we simulate planet-gas interactions in debris disks with the help of hydrodynamical simulations. This allows us to model the opening of the gap than with the help of analytical 1D models and to easily produce synthetic images. This hydrodynamical modeling requires a good estimate of the disk's properties. Observations as listed in Table \ref{tab:known_disks} are essential, of course; however, there are still some properties that are difficult to constrain. 

\subsubsection{Disk turbulence and viscosity}

The turbulence of the gas is one of the big unknowns of the system. In \cite{Kral2016a}, the authors try to constrain the level of turbulence in the gas of the $\beta$ Pictoris system for the first time. Thanks to a particularly efficient ionization of the second generation gas, they find that the turbulent $\alpha$  parameter can be as high as 0.1, which is consistent with an MRI-driven turbulence \citep{Kral2016b,Marino2020}, but it can also range over multiple orders of magnitude. This uncertainty rises from the difficulty in observationally constraining other parameters such as the accretion rate of the gas. 

The $\alpha$ parameter is also poorly constrained in protoplanetary disks, even though recent studies show that the disks are probably of low turbulence \citep{Dullemond2020,Pascucci2023}. Therefore, regarding these uncertainties, we decided to investigate the formation of gaps and kinks in debris disks with intermediate turbulence \citep[i.e., $\alpha = 10^{-3}$,][]{Cui2024} . This choice influences how a gap is opened; in a disk hosting a planet of a given mass and distance from the star, the gap opened by the planet is deeper for disks with lower viscosity and therefore lower $\alpha$ \citep{Crida2006,Fung2014,Kanagawa2015}. Consequently, if $\alpha$ is higher than in our study, either kinks become observable or one needs larger planetary masses to reproduce the same gap depth. However, what is important in our study is the capacity of the instruments (here ALMA) to detect the gas located outside of the gap but not inside. Therefore, our study provides a first approach to determine the observability of planetary substructures in debris disks. Future extensive parameter studies are required to precisely estimate the planet characteristics needed to observe substructures in specific targets. 

\subsubsection{Disk temperature}

The temperature of the gas in debris disks is slightly better constrained from observations. As mentioned in Sect. \ref{sec:numericalsetup}, the temperature that we used in our simulations is consistent with what is expected from the release of second-generation gas and observations \citep{Kral2019}. However, our hydrodynamical simulations are locally isothermal, which can influence the opening of the gap \citep{Les2015}. Therefore, it is obvious that more precise, 3D hydrodynamical simulations taking into account a precise vertical distribution of the temperature in the specific framework of debris disks are needed to correctly simulate gap opening. However, this would be computationally expensive and it would require a precise estimate for the disk temperature structure from the observations, which is still debated and difficult to obtain. Therefore, our 2D isothermal setup is a great first step in the modeling of the planet-gas interactions in debris disks.

\subsubsection{Disk radial extent and distance}
\label{subsubsec:discuss_diskdist}

The radial extent of disks investigated in this study is consistent with the largest disks observed nowadays. The total range of the gaseous disk can impact its observability on the same basis as its distance from Earth. Regarding the clear detection of the planetary gaps, the disk needs to be sufficiently larger than the gap width to be detected outside of the gap. This requires the correct combination of gap width, gap depth, beam size, and flux emitted from the gas outside of the gap in one beam. 

The detection of planetary kinks requires different conditions regarding the disk's size and distance. Here, the important aspect relies on the gas located close to the planet's orbit. Therefore, compared to the conditions required to observe gaps, the disk can be less extended. However, the gas needs to be luminous close to the planet, and the detection of the kink requires a high angular resolution. The size of the kink can be estimated with different tools \citep[e.g.,][]{Bollati2021,Izquierdo2021,Izquierdo2023}. In general, it is smaller than the gap width for a planet of a given mass and semi-major axis. Difficulties in detecting kinks are therefore similar to the ones encountered for their detection in protoplanetary disks.

In Appendix \ref{appendix_distances}, we show the images of the disks located at different distances from Earth (i.e., at 100 pc and 130 pc). From these images, we clearly see that the observability of the gap depends on the ability to resolve the gap width. We find that the wider gaps are still observable even at 130 pc. Moreover, the inner and outer parts of the disk need to be luminous and at least as wide as the gap. 

\subsection{Correctly determining $\Sigma_{\rm gap}$}
\label{sec:discuss_sigmagap}

In Sect. \ref{sec:gapobserv}, we show that it is possible to determine the integrated flux emitted by a parcel of gas located in the gap from the surface density present inside the gap $\Sigma_{\rm gap}$. Several studies derived criteria to estimate the gap depth from the disk's parameters. \cite{Kanagawa2015} found that the depth of the gap can be written as
\begin{equation}
    \frac{\Sigma_{\rm gap}}{\Sigma_0} = \frac{1}{1 + 0.04K}
    \label{eq4}
,\end{equation}

\noindent where $K = q^2h^{-5}\alpha^{-1}$; with $q$ being the planet-to-star-mass ratio and $\Sigma_0$ the initial surface density at the planet's position. This criterion is valid for $K \leqslant 10^{4}$. It has been corrected in different frameworks; for example, in \cite{Yun2019} they adapted this criterion to nonuniform disks (i.e., a power-law initial surface-density profile with an exponential cutoff), finding that $\Sigma_{\rm gap}/\Sigma_0 = 1/( 1 + 0.046K)$. In \cite{Pichierri2023}, they corrected Eq. \ref{eq4} to correctly describe the partial gap opening of planets in low-viscosity disks. They found that $\Sigma_{\rm gap}/\Sigma_0 = 1/( 1 + 0.04 \Tilde{K})$ with $\Tilde{K} = 3.93q^{2.3}h^{-6.14}\alpha^{-0.66}$. 

\begin{figure}[t]
   \centering   
   \includegraphics[scale=0.37]{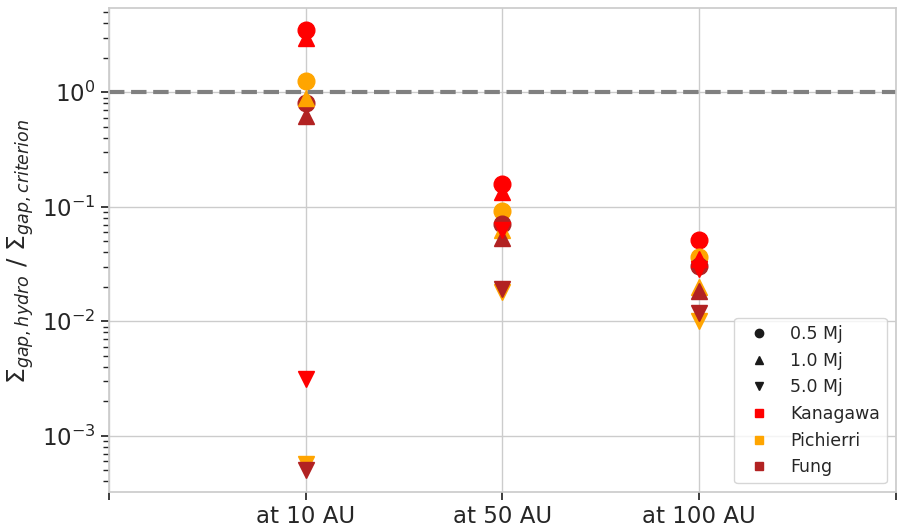}
   \caption{Ratio between averaged surface density inside the gaps in our hydrodynamical simulations $\Sigma_{\rm gap,hydro}$ and the surface density estimated from different criteria $\Sigma_{\rm gap, criterion}$. In red are the estimates from \cite{Kanagawa2015}, in orange those from \cite{Pichierri2023} and in brown those from \cite{Fung2014}. Circles represent the cases where $m_p = 0.5 \; \rm M_J$, upward triangles are for $m_p = 1 \; \rm M_J,$ and downward triangles are for $m_p = 5 \; \rm M_J$. We see that the different criteria fail to reproduce our gap depth the further away the planet is located. For the planet located at 10 AU, the criteria fail to reproduce the $m_p = 5 \; \rm M_J$ case as the planets become eccentric, changing their gap opening. }
   \label{gap_comp}
\end{figure}

\cite{Fung2014} also derived a formula describing the gap depth as a function of the planet and disk characteristics. They found that for planetary masses ranging from $10^{-4} \leqslant q \leqslant 5\times10^{-3}$, for $10^{-3} \leqslant \alpha \leqslant 10^{-1}$ and  $0.04 \leqslant h \leqslant 0.1$,
\begin{equation}
    \frac{\Sigma_{\rm gap}}{\Sigma_0} = 0.14 \left(\frac{q}{10^{-3}}\right)^{-2.16} \left(\frac{\alpha}{10^{-2}}\right)^{1.41} \left(\frac{h}{0.05}\right)^{6.61}
    \label{eq5}
.\end{equation}

A comparison between these different criteria can be found in \cite{Bergez2020}, where they also state that taking planetary gas accretion into account can impact the way planets open gaps, especially at low viscosity. 

In our study, $\Sigma_{\rm gap}$ was directly determined from our hydrodynamical simulations in order to be consistent with our synthetic images (see Appendix \ref{appendix_hydro}). As shown in Fig. \ref{gap_comp}, we notice a difference in gap depth between our simulations and the different criteria cited above. This can originate from the different frameworks investigated. In our case, we are at pretty low viscosity $\nu,$ as $\alpha = 10^{-3}$ and $h = 0.01(r/1AU)^{0.25}$. Therefore, our gap depth cannot be compared to that of \cite{Fung2014}. Regarding the criterion of \cite{Kanagawa2015}, we are in a low-viscosity disk; therefore, we need to compare to \cite{Pichierri2023}. However, the majority of our planets are more massive than those in their study, and therefore they do not open partial gap, but rather deep gaps. A different behavior us thus to be expected.

Determining the correct depth of the gap can be tricky depending on the characteristics of both the planet and the disk gas. However, we estimate that the listed gap-opening criteria can still give an interesting first estimate of the gap depth to predict wether the gap would be observable or not. In Fig. \ref{gap_comp}, we quantify that in the worst case scenario, the gap depths are off by a bit more than one order of magnitude in the low-mass cases, where the planets are not eccentric at all. Future complex hydrodynamical simulations taking into account 3D effects in radiative disks will help derive a robust gap-opening criterion.

\subsection{Influence of planet characteristics}
\label{sec:discuss_parameters}

Many different parameters can influence the observability of the planet-gas interactions: the disk characteristics, the observational configuration (see Sect. \ref{sec:discuss_modelassump}), and the planet characteristics. The planet's mass and semi-major axis are the two main properties that influence the observability of the substructures. They dictate how deep the gap is in a given disk (see Sect. \ref{sec:discuss_sigmagap}) and how strong the spiral arms are. Consequently, a less massive planet produces a shallower and narrower gap as well as weaker spiral arms \citep{GoodmanandRafikov2001}. 

As mentioned in the previous section, the depth of the gap also highly depends on the properties of the gas (specifically its viscosity $\nu \propto \alpha h^2$). Therefore, a Saturn-mass planet will easily produce a deep gap in a low-turbulence, cold disk, but it will hardly carve a gap in a hot, highly turbulent one. As the gas in debris disks is estimated to be richer in heavy elements than protoplanetary disks, in general they present lower aspect ratios than in protoplanetary disks, making it easy for lower mass planets to open deep gaps. Depending on the $\alpha$ parameter, one can therefore expect planet masses as low as those of Saturn or even Neptune to carve deep gaps; assuming that $h = 0.02$ at the planet position, in an $\alpha = 10^{-3}$ disk, \cite{Kanagawa2015} estimated that a Neptune-mass planet produces a gap depth of $\Sigma_{\rm gap}/\Sigma_0 = 0.03$, which would definitely produce an observable gap \citep[a deep gap is estimated to be opened when $\Sigma_{\rm gap}/\Sigma_0 < 0.1$][]{Crida2006}. However, in Neptune's case, its gap would be very narrow: at 10 AU, $4r_H$ = 1.03 AU; at 50 AU, $4r_H$ = 5.15 AU; and at 100 AU, $4r_H$ = 10.32 AU. Compared to the gap widths investigated in this work (see Table \ref{tab:gap_widths}), we see that it can be difficult to observe such a gap in the C-6 configuration in Band 6. However, one can still find a configuration where a Neptune gap would be observable (maybe when it is located deep in a cold disk, or using another configuration or band; see Sect. \ref{sec:discuss_gascompo}).

Regarding the observability of the kinks, they will only be observable for the most massive planets (i.e., $m_p > 1 \; \rm M_J$), located at large semi-major axes. As discussed in Sect. \ref{subsubsec:discuss_diskdist}, their observability is as limited as in protoplanetary disks. 

\subsection{Observations in other species than $\rm ^{12}CO$}
\label{sec:discuss_gascompo}

In this study, we focused on the CO(J = 2-1) emission of the gas present in our debris disks. However, other CO lines (e.g., J = 1-0 or J = 3-2) and other components have been observed in some disks (e.g., CI or OI; see Table \ref{appendix_knowndisks}). In order for CO to survive over large timescales in debris disks, it needs to be protected from the external irradiation that would photodissociate the molecule \citep[in about 120 yr when considering photons from the interstellar radiation field,][]{Visser2009}. Therefore, the presence of CO is expected to be consistent with CI gas, shielding the CO in disks more massive than $\gtrsim 10^{-3} \; \rm M_\oplus$ \citep{Kral2019,Marino2020,Cataldi2023}. The planets will interact with all the different components of the gas disk, meaning that the planetary gaps can also be observed in ALMA Band 8, probing the CI emission. We note that disks that are not shielded should not be targeted in CO but rather in neutral carbon. This is because CO would be photodissociated before having time to viscously spread when no shielding by carbon operates, whereas carbon still viscously spreads inwards and outwards.

Moreover, different CO isotopologues can be used to probe the disks at different scales. As in general $\rm ^{12}CO$ is brighter than, for example, $\rm ^{13}CO$ and $\rm C^{18}O$, one can use their different intensity to look for the signature of the gap. In the massive disks where kinks might be observed, one can see that the same system in these isotopologues may have less bright emission in the gap and validate our flux criterion (Eq. \ref{eq_fluxcriterion}).

More generally, our criterion Eq. \ref{eq_fluxcriterion} can be adapted to any kind of emission line; what matters is the capacity to resolve the gap and to detect the gas outside of the gap. Once one has the properties of some gas distribution (radial extent; estimations for its temperature and viscosity), then our analysis can be applied.

\subsection{Gaps in the dust versus gaps in the gas}
\label{sec:discuss_gapdust}

Debris disks are often seen as broad disks, sometimes hosting multiple gaps in them \citep[e.g.,][]{Su2013, Marino2018, Esposito2020,Faramaz2021}. The structure of the planetesimal and dust distributions can be influenced by planets; the analysis of the steepness of the inner edge of the disk is consistent with the presence of planets that are too cold to be imaged directly, deriving a first estimate of the planet's characteristics \citep[e.g.,][]{Wyatt2008,Pearce2022,ImazBlanco2023,Marshall2023}. 

Our approach using the gas in debris disks brings complementary information compared to these analyses from dust observations for different reasons: i) in some disks, the gas is more radially extended than the dust/planetesimal disk (e.g., in HD138813 and HD156623), meaning that we can probe a different exoplanet population that is too far to significantly perturb the planetesimal disks; ii) the indirect detections of a planet from the dust or gas observations are independent, meaning that the planet characteristics can be derived independently; iii) when observing a dust gap, there are still uncertainties on the presence of a planet within the gap because the latter could be located further inwards, as shown in \cite{Pearce2015}, owing to secular effects of the planets. Moreover, migrating planets can affect the dust-gap width, but those effects would have less impact on the gas distribution, which would be much more reliable. A multiwavelength study could therefore really improve our understanding on planet-disk interactions, either by confirming gap-opening processes or by improving their estimation (gap-opening criterion). Such observations could also be extended to gap formation during the protoplanetary-disk phase.

\vspace{3cm}

\section{Conclusions}
\label{sec:conclusion}

In this paper, we derived synthetic images of the CO(J = 2-1) line emission of the gas present in typical debris disks. The gas is perturbed by a single planet with varying mass and semi-major axis. After deriving the perturbed profiles with the help of 2D hydrodynamical simulations, realistic synthetic images are produced from a radiative transfer code and the ALMA observing tool. Our conclusions are as follows.

\begin{enumerate}
    \item Planet-gas interactions are also observable in debris disks. While it has been shown in previous studies that planet-gas interactions can produce observable substructures in protoplanetary disks (e.g., kinks), we show here that these kinds of interactions are also visible in older debris disks. These observations can lead to the indirect discovery of new exoplanets, at an intermediate stage of their evolution (in old systems compared to the protoplanetary disk phase). Moreover, this can help us discover cold giant planets that are not luminous enough to be detected via direct imaging.
    
    \item More importantly, unlike in protoplanetary disks, we are able to observe gaseous planetary gaps in debris disks. Thanks to their low surface density, the emission of the gas located in the gap can be sufficiently low to remain below ALMA's sensitivity threshold. We derived a criterion (Eq. \ref{eq_fluxcriterion}) able to estimate whether gaps are observable depending on the planet, disk, and observational configuration characteristics.

    \item Depending on the gap depth and disk mass, we are able to detect either gaps or kinks: if there is still a sufficient amount of gas around the planet (e.g., in the most massive gaseous debris disks), kinks are observable for the most massive planets (i.e., $m_p > 1 \; \rm M_J$) and the gaps are too luminous to be detected. However, for a more moderate disk mass (i.e., $m_{disk} \approx 10^{-3} \; \rm M_\oplus$), the same planet will produce an observable gap, but its kink will no longer be visible. Depending on the gap depth and widths, planets as small as Saturn might produce observable gaps.
        
    \item By applying our method to a known debris disk (HD 138813), we show that we can realistically already detect such gaps if they exist. Thanks to our large parameter space study (planet masses, locations, and disk masses and inclinations), we estimate that several debris disks known to host gas are ideal candidates to look for gaseous planetary gaps.
    
\end{enumerate}

This study may have several important impacts on our understanding of gas in debris disks. First, it might lead to a new way to indirectly detect exoplanets at an intermediate stage of their formation, where the majority of the primordial gas is gone but the planetesimals' belts are still massive. Secondly, as gaps are only observable in the dust in protoplanetary disks, being able to compare the shape of the gap edges in both the dust and the gas in debris disks can help constrain gap formation models. Future observations at high angular resolutions of the gas in debris disks are to be combined with further studies modeling the gas evolution in the presence of planets.

\begin{acknowledgements}
      C. Bergez-Casalou thanks the PSL Fellowship for their financial support. C. Bergez-Casalou also thanks L. Matrà and S. Marino for the interesting discussions on this project, which helped improve its development.
\end{acknowledgements}

\bibliographystyle{aa}
\bibliography{biblio}

\begin{thebibliography}{89}
\expandafter\ifx\csname natexlab\endcsname\relax\def\natexlab#1{#1}\fi

\bibitem[{{Armitage}(2022)}]{ArmitageLN2022}
{Armitage}, P.~J. 2022, arXiv e-prints, arXiv:2201.07262

\bibitem[{{Barraza-Alfaro} {et~al.}(2024){Barraza-Alfaro}, {Flock}, \& {Henning}}]{Barraza2024}
{Barraza-Alfaro}, M., {Flock}, M., \& {Henning}, T. 2024, \aap, 683, A16

\bibitem[{{Ben{\'\i}tez-Llambay} \& {Masset}(2016)}]{Benitez-Llambay2016}
{Ben{\'\i}tez-Llambay}, P. \& {Masset}, F.~S. 2016, \apjs, 223, 11

\bibitem[{{Bergez-Casalou} {et~al.}(2022){Bergez-Casalou}, {Bitsch}, {Kurtovic}, \& {Pinilla}}]{Bergez2022}
{Bergez-Casalou}, C., {Bitsch}, B., {Kurtovic}, N.~T., \& {Pinilla}, P. 2022, \aap, 659, A6

\bibitem[{{Bergez-Casalou} {et~al.}(2020){Bergez-Casalou}, {Bitsch}, {Pierens}, {Crida}, \& {Raymond}}]{Bergez2020}
{Bergez-Casalou}, C., {Bitsch}, B., {Pierens}, A., {Crida}, A., \& {Raymond}, S.~N. 2020, \aap, 643, A133

\bibitem[{{Bitsch} {et~al.}(2013){Bitsch}, {Crida}, {Libert}, \& {Lega}}]{Bitsch2013eccentricity}
{Bitsch}, B., {Crida}, A., {Libert}, A.~S., \& {Lega}, E. 2013, \aap, 555, A124

\bibitem[{{Bitsch} \& {Kley}(2010)}]{Bitsch2010}
{Bitsch}, B. \& {Kley}, W. 2010, \aap, 523, A30

\bibitem[{{Bollati} {et~al.}(2021){Bollati}, {Lodato}, {Price}, \& {Pinte}}]{Bollati2021}
{Bollati}, F., {Lodato}, G., {Price}, D.~J., \& {Pinte}, C. 2021, \mnras, 504, 5444

\bibitem[{{Bonsor} {et~al.}(2023){Bonsor}, {Wyatt}, {Marino}, {Davidsson}, {Kral}, \& {Thebault}}]{Bonsor2023}
{Bonsor}, A., {Wyatt}, M.~C., {Marino}, S., {et~al.} 2023, \mnras, 526, 3115

\bibitem[{{Booth} {et~al.}(2019){Booth}, {Matr{\`a}}, {Su}, {Kral}, {Hales}, {Dent}, {Hughes}, {MacGregor}, {L{\"o}hne}, \& {Wilner}}]{Booth2019}
{Booth}, M., {Matr{\`a}}, L., {Su}, K. Y.~L., {et~al.} 2019, \mnras, 482, 3443

\bibitem[{{Brandeker} {et~al.}(2016){Brandeker}, {Cataldi}, {Olofsson}, {Vandenbussche}, {Acke}, {Barlow}, {Blommaert}, {Cohen}, {Dent}, {Dominik}, {Di Francesco}, {Fridlund}, {Gear}, {Glauser}, {Greaves}, {Harvey}, {Heras}, {Hogerheijde}, {Holland}, {Huygen}, {Ivison}, {Leeks}, {Lim}, {Liseau}, {Matthews}, {Pantin}, {Pilbratt}, {Royer}, {Sibthorpe}, {Waelkens}, \& {Walker}}]{Brandeker2016}
{Brandeker}, A., {Cataldi}, G., {Olofsson}, G., {et~al.} 2016, \aap, 591, A27

\bibitem[{{CASA Team} {et~al.}(2022){CASA Team}, {Bean}, {Bhatnagar}, {Castro}, {Donovan Meyer}, {Emonts}, {Garcia}, {Garwood}, {Golap}, {Gonzalez Villalba}, {Harris}, {Hayashi}, {Hoskins}, {Hsieh}, {Jagannathan}, {Kawasaki}, {Keimpema}, {Kettenis}, {Lopez}, {Marvil}, {Masters}, {McNichols}, {Mehringer}, {Miel}, {Moellenbrock}, {Montesino}, {Nakazato}, {Ott}, {Petry}, {Pokorny}, {Raba}, {Rau}, {Schiebel}, {Schweighart}, {Sekhar}, {Shimada}, {Small}, {Steeb}, {Sugimoto}, {Suoranta}, {Tsutsumi}, {van Bemmel}, {Verkouter}, {Wells}, {Xiong}, {Szomoru}, {Griffith}, {Glendenning}, \& {Kern}}]{CASA}
{CASA Team}, {Bean}, B., {Bhatnagar}, S., {et~al.} 2022, \pasp, 134, 114501

\bibitem[{{Cataldi} {et~al.}(2023){Cataldi}, {Aikawa}, {Iwasaki}, {Marino}, {Brandeker}, {Hales}, {Henning}, {Higuchi}, {Hughes}, {Janson}, {Kral}, {Matr{\`a}}, {Mo{\'o}r}, {Olofsson}, {Redfield}, \& {Roberge}}]{Cataldi2023}
{Cataldi}, G., {Aikawa}, Y., {Iwasaki}, K., {et~al.} 2023, \apj, 951, 111

\bibitem[{{Cataldi} {et~al.}(2014){Cataldi}, {Brandeker}, {Olofsson}, {Larsson}, {Liseau}, {Blommaert}, {Fridlund}, {Ivison}, {Pantin}, {Sibthorpe}, {Vandenbussche}, \& {Wu}}]{Cataldi2014}
{Cataldi}, G., {Brandeker}, A., {Olofsson}, G., {et~al.} 2014, \aap, 563, A66

\bibitem[{{Cataldi} {et~al.}(2020){Cataldi}, {Wu}, {Brandeker}, {Ohashi}, {Mo{\'o}r}, {Olofsson}, {{\'A}brah{\'a}m}, {Asensio-Torres}, {Cavallius}, {Dent}, {Grady}, {Henning}, {Higuchi}, {Hughes}, {Janson}, {Kamp}, {K{\'o}sp{\'a}l}, {Redfield}, {Roberge}, {Weinberger}, \& {Welsh}}]{Cataldi2020}
{Cataldi}, G., {Wu}, Y., {Brandeker}, A., {et~al.} 2020, \apj, 892, 99

\bibitem[{{Crida} {et~al.}(2006){Crida}, {Morbidelli}, \& {Masset}}]{Crida2006}
{Crida}, A., {Morbidelli}, A., \& {Masset}, F. 2006, \icarus, 181, 587

\bibitem[{{Cui} {et~al.}(2024){Cui}, {Marino}, {Kral}, \& {Latter}}]{Cui2024}
{Cui}, C., {Marino}, S., {Kral}, Q., \& {Latter}, H. 2024, \mnras, 530, 1766

\bibitem[{{Dent} {et~al.}(2014){Dent}, {Wyatt}, {Roberge}, {Augereau}, {Casassus}, {Corder}, {Greaves}, {de Gregorio-Monsalvo}, {Hales}, {Jackson}, {Hughes}, {Lagrange}, {Matthews}, \& {Wilner}}]{Dent2014}
{Dent}, W.~R.~F., {Wyatt}, M.~C., {Roberge}, A., {et~al.} 2014, Science, 343, 1490

\bibitem[{{Di Folco} {et~al.}(2020){Di Folco}, {P{\'e}ricaud}, {Dutrey}, {Augereau}, {Chapillon}, {Guilloteau}, {Pi{\'e}tu}, \& {Boccaletti}}]{DiFolco2020}
{Di Folco}, E., {P{\'e}ricaud}, J., {Dutrey}, A., {et~al.} 2020, \aap, 635, A94

\bibitem[{{Donaldson} {et~al.}(2013){Donaldson}, {Lebreton}, {Roberge}, {Augereau}, \& {Krivov}}]{Donaldson2013}
{Donaldson}, J.~K., {Lebreton}, J., {Roberge}, A., {Augereau}, J.~C., \& {Krivov}, A.~V. 2013, \apj, 772, 17

\bibitem[{{Dullemond} {et~al.}(2020){Dullemond}, {Isella}, {Andrews}, {Skobleva}, \& {Dzyurkevich}}]{Dullemond2020}
{Dullemond}, C.~P., {Isella}, A., {Andrews}, S.~M., {Skobleva}, I., \& {Dzyurkevich}, N. 2020, \aap, 633, A137

\bibitem[{{Dullemond} {et~al.}(2012){Dullemond}, {Juhasz}, {Pohl}, {Sereshti}, {Shetty}, {Peters}, {Commercon}, \& {Flock}}]{Dullemond2012}
{Dullemond}, C.~P., {Juhasz}, A., {Pohl}, A., {et~al.} 2012, {RADMC-3D: A multi-purpose radiative transfer tool}

\bibitem[{{Esposito} {et~al.}(2020){Esposito}, {Kalas}, {Fitzgerald}, {Millar-Blanchaer}, {Duch{\^e}ne}, {Patience}, {Hom}, {Perrin}, {De Rosa}, {Chiang}, {Czekala}, {Macintosh}, {Graham}, {Ansdell}, {Arriaga}, {Bruzzone}, {Bulger}, {Chen}, {Cotten}, {Dong}, {Draper}, {Follette}, {Hung}, {Lopez}, {Matthews}, {Mazoyer}, {Metchev}, {Rameau}, {Ren}, {Rice}, {Song}, {Stahl}, {Wang}, {Wolff}, {Zuckerman}, {Ammons}, {Bailey}, {Barman}, {Chilcote}, {Doyon}, {Gerard}, {Goodsell}, {Greenbaum}, {Hibon}, {Hinkley}, {Ingraham}, {Konopacky}, {Maire}, {Marchis}, {Marley}, {Marois}, {Nielsen}, {Oppenheimer}, {Palmer}, {Poyneer}, {Pueyo}, {Rajan}, {Rantakyr{\"o}}, {Ruffio}, {Savransky}, {Schneider}, {Sivaramakrishnan}, {Soummer}, {Thomas}, \& {Ward-Duong}}]{Esposito2020}
{Esposito}, T.~M., {Kalas}, P., {Fitzgerald}, M.~P., {et~al.} 2020, \aj, 160, 24

\bibitem[{{Faramaz} {et~al.}(2021){Faramaz}, {Marino}, {Booth}, {Matr{\`a}}, {Mamajek}, {Bryden}, {Stapelfeldt}, {Casassus}, {Cuadra}, {Hales}, \& {Zurlo}}]{Faramaz2021}
{Faramaz}, V., {Marino}, S., {Booth}, M., {et~al.} 2021, \aj, 161, 271

\bibitem[{{Fulton} {et~al.}(2021){Fulton}, {Rosenthal}, {Hirsch}, {Isaacson}, {Howard}, {Dedrick}, {Sherstyuk}, {Blunt}, {Petigura}, {Knutson}, {Behmard}, {Chontos}, {Crepp}, {Crossfield}, {Dalba}, {Fischer}, {Henry}, {Kane}, {Kosiarek}, {Marcy}, {Rubenzahl}, {Weiss}, \& {Wright}}]{Fulton2021}
{Fulton}, B.~J., {Rosenthal}, L.~J., {Hirsch}, L.~A., {et~al.} 2021, \apjs, 255, 14

\bibitem[{{Fung} {et~al.}(2014){Fung}, {Shi}, \& {Chiang}}]{Fung2014}
{Fung}, J., {Shi}, J.-M., \& {Chiang}, E. 2014, \apj, 782, 88

\bibitem[{{Gaia Collaboration} {et~al.}(2023){Gaia Collaboration}, {Vallenari}, {Brown}, {Prusti}, {de Bruijne}, {Arenou}, {Babusiaux}, {Biermann}, {Creevey}, {Ducourant}, {Evans}, {Eyer}, {Guerra}, {Hutton}, {Jordi}, {Klioner}, {Lammers}, {Lindegren}, {Luri}, {Mignard}, {Panem}, {Pourbaix}, {Randich}, {Sartoretti}, {Soubiran}, {Tanga}, {Walton}, {Bailer-Jones}, {Bastian}, {Drimmel}, {Jansen}, {Katz}, {Lattanzi}, {van Leeuwen}, {Bakker}, {Cacciari}, {Casta{\~n}eda}, {De Angeli}, {Fabricius}, {Fouesneau}, {Fr{\'e}mat}, {Galluccio}, {Guerrier}, {Heiter}, {Masana}, {Messineo}, {Mowlavi}, {Nicolas}, {Nienartowicz}, {Pailler}, {Panuzzo}, {Riclet}, {Roux}, {Seabroke}, {Sordo}, {Th{\'e}venin}, {Gracia-Abril}, {Portell}, {Teyssier}, {Altmann}, {Andrae}, {Audard}, {Bellas-Velidis}, {Benson}, {Berthier}, {Blomme}, {Burgess}, {Busonero}, {Busso}, {C{\'a}novas}, {Carry}, {Cellino}, {Cheek}, {Clementini}, {Damerdji}, {Davidson}, {de Teodoro}, {Nu{\~n}ez Campos}, {Delchambre}, {Dell'Oro}, {Esquej},
  {Fern{\'a}ndez-Hern{\'a}ndez}, {Fraile}, {Garabato}, {Garc{\'\i}a-Lario}, {Gosset}, {Haigron}, {Halbwachs}, {Hambly}, {Harrison}, {Hern{\'a}ndez}, {Hestroffer}, {Hodgkin}, {Holl}, {Jan{\ss}en}, {Jevardat de Fombelle}, {Jordan}, {Krone-Martins}, {Lanzafame}, {L{\"o}ffler}, {Marchal}, {Marrese}, {Moitinho}, {Muinonen}, {Osborne}, {Pancino}, {Pauwels}, {Recio-Blanco}, {Reyl{\'e}}, {Riello}, {Rimoldini}, {Roegiers}, {Rybizki}, {Sarro}, {Siopis}, {Smith}, {Sozzetti}, {Utrilla}, {van Leeuwen}, {Abbas}, {{\'A}brah{\'a}m}, {Abreu Aramburu}, {Aerts}, {Aguado}, {Ajaj}, {Aldea-Montero}, {Altavilla}, {{\'A}lvarez}, {Alves}, {Anders}, {Anderson}, {Anglada Varela}, {Antoja}, {Baines}, {Baker}, {Balaguer-N{\'u}{\~n}ez}, {Balbinot}, {Balog}, {Barache}, {Barbato}, {Barros}, {Barstow}, {Bartolom{\'e}}, {Bassilana}, {Bauchet}, {Becciani}, {Bellazzini}, {Berihuete}, {Bernet}, {Bertone}, {Bianchi}, {Binnenfeld}, {Blanco-Cuaresma}, {Blazere}, {Boch}, {Bombrun}, {Bossini}, {Bouquillon}, {Bragaglia}, {Bramante}, {Breedt},
  {Bressan}, {Brouillet}, {Brugaletta}, {Bucciarelli}, {Burlacu}, {Butkevich}, {Buzzi}, {Caffau}, {Cancelliere}, {Cantat-Gaudin}, {Carballo}, {Carlucci}, {Carnerero}, {Carrasco}, {Casamiquela}, {Castellani}, {Castro-Ginard}, {Chaoul}, {Charlot}, {Chemin}, {Chiaramida}, {Chiavassa}, {Chornay}, {Comoretto}, {Contursi}, {Cooper}, {Cornez}, {Cowell}, {Crifo}, {Cropper}, {Crosta}, {Crowley}, {Dafonte}, {Dapergolas}, {David}, {David}, {de Laverny}, {De Luise}, {De March}, {De Ridder}, {de Souza}, {de Torres}, {del Peloso}, {del Pozo}, {Delbo}, {Delgado}, {Delisle}, {Demouchy}, {Dharmawardena}, {Di Matteo}, {Diakite}, {Diener}, {Distefano}, {Dolding}, {Edvardsson}, {Enke}, {Fabre}, {Fabrizio}, {Faigler}, {Fedorets}, {Fernique}, {Fienga}, {Figueras}, {Fournier}, {Fouron}, {Fragkoudi}, {Gai}, {Garcia-Gutierrez}, {Garcia-Reinaldos}, {Garc{\'\i}a-Torres}, {Garofalo}, {Gavel}, {Gavras}, {Gerlach}, {Geyer}, {Giacobbe}, {Gilmore}, {Girona}, {Giuffrida}, {Gomel}, {Gomez}, {Gonz{\'a}lez-N{\'u}{\~n}ez},
  {Gonz{\'a}lez-Santamar{\'\i}a}, {Gonz{\'a}lez-Vidal}, {Granvik}, {Guillout}, {Guiraud}, {Guti{\'e}rrez-S{\'a}nchez}, {Guy}, {Hatzidimitriou}, {Hauser}, {Haywood}, {Helmer}, {Helmi}, {Sarmiento}, {Hidalgo}, {Hilger}, {H{\l}adczuk}, {Hobbs}, {Holland}, {Huckle}, {Jardine}, {Jasniewicz}, {Jean-Antoine Piccolo}, {Jim{\'e}nez-Arranz}, {Jorissen}, {Juaristi Campillo}, {Julbe}, {Karbevska}, {Kervella}, {Khanna}, {Kontizas}, {Kordopatis}, {Korn}, {K{\'o}sp{\'a}l}, {Kostrzewa-Rutkowska}, {Kruszy{\'n}ska}, {Kun}, {Laizeau}, {Lambert}, {Lanza}, {Lasne}, {Le Campion}, {Lebreton}, {Lebzelter}, {Leccia}, {Leclerc}, {Lecoeur-Taibi}, {Liao}, {Licata}, {Lindstr{\o}m}, {Lister}, {Livanou}, {Lobel}, {Lorca}, {Loup}, {Madrero Pardo}, {Magdaleno Romeo}, {Managau}, {Mann}, {Manteiga}, {Marchant}, {Marconi}, {Marcos}, {Marcos Santos}, {Mar{\'\i}n Pina}, {Marinoni}, {Marocco}, {Marshall}, {Martin Polo}, {Mart{\'\i}n-Fleitas}, {Marton}, {Mary}, {Masip}, {Massari}, {Mastrobuono-Battisti}, {Mazeh}, {McMillan}, {Messina}, {Michalik},
  {Millar}, {Mints}, {Molina}, {Molinaro}, {Moln{\'a}r}, {Monari}, {Mongui{\'o}}, {Montegriffo}, {Montero}, {Mor}, {Mora}, {Morbidelli}, {Morel}, {Morris}, {Muraveva}, {Murphy}, {Musella}, {Nagy}, {Noval}, {Oca{\~n}a}, {Ogden}, {Ordenovic}, {Osinde}, {Pagani}, {Pagano}, {Palaversa}, {Palicio}, {Pallas-Quintela}, {Panahi}, {Payne-Wardenaar}, {Pe{\~n}alosa Esteller}, {Penttil{\"a}}, {Pichon}, {Piersimoni}, {Pineau}, {Plachy}, {Plum}, {Poggio}, {Pr{\v{s}}a}, {Pulone}, {Racero}, {Ragaini}, {Rainer}, {Raiteri}, {Rambaux}, {Ramos}, {Ramos-Lerate}, {Re Fiorentin}, {Regibo}, {Richards}, {Rios Diaz}, {Ripepi}, {Riva}, {Rix}, {Rixon}, {Robichon}, {Robin}, {Robin}, {Roelens}, {Rogues}, {Rohrbasser}, {Romero-G{\'o}mez}, {Rowell}, {Royer}, {Ruz Mieres}, {Rybicki}, {Sadowski}, {S{\'a}ez N{\'u}{\~n}ez}, {Sagrist{\`a} Sell{\'e}s}, {Sahlmann}, {Salguero}, {Samaras}, {Sanchez Gimenez}, {Sanna}, {Santove{\~n}a}, {Sarasso}, {Schultheis}, {Sciacca}, {Segol}, {Segovia}, {S{\'e}gransan}, {Semeux}, {Shahaf}, {Siddiqui}, {Siebert},
  {Siltala}, {Silvelo}, {Slezak}, {Slezak}, {Smart}, {Snaith}, {Solano}, {Solitro}, {Souami}, {Souchay}, {Spagna}, {Spina}, {Spoto}, {Steele}, {Steidelm{\"u}ller}, {Stephenson}, {S{\"u}veges}, {Surdej}, {Szabados}, {Szegedi-Elek}, {Taris}, {Taylor}, {Teixeira}, {Tolomei}, {Tonello}, {Torra}, {Torra}, {Torralba Elipe}, {Trabucchi}, {Tsounis}, {Turon}, {Ulla}, {Unger}, {Vaillant}, {van Dillen}, {van Reeven}, {Vanel}, {Vecchiato}, {Viala}, {Vicente}, {Voutsinas}, {Weiler}, {Wevers}, {Wyrzykowski}, {Yoldas}, {Yvard}, {Zhao}, {Zorec}, {Zucker}, \& {Zwitter}}]{GaiaDR3}
{Gaia Collaboration}, {Vallenari}, A., {Brown}, A.~G.~A., {et~al.} 2023, \aap, 674, A1

\bibitem[{{Goodman} \& {Rafikov}(2001)}]{GoodmanandRafikov2001}
{Goodman}, J. \& {Rafikov}, R.~R. 2001, \apj, 552, 793

\bibitem[{{Gyeol Yun} {et~al.}(2019){Gyeol Yun}, {Kim}, {Bae}, \& {Han}}]{Yun2019}
{Gyeol Yun}, H., {Kim}, W.-T., {Bae}, J., \& {Han}, C. 2019, \apj, 884, 142

\bibitem[{{Hales} {et~al.}(2019){Hales}, {Gorti}, {Carpenter}, {Hughes}, \& {Flaherty}}]{Hales2019}
{Hales}, A.~S., {Gorti}, U., {Carpenter}, J.~M., {Hughes}, M., \& {Flaherty}, K. 2019, \apj, 878, 113

\bibitem[{{Hales} {et~al.}(2022){Hales}, {Marino}, {Sheehan}, {Ulloa}, {P{\'e}rez}, {Matr{\`a}}, {Kral}, {Wyatt}, {Dent}, \& {Carpenter}}]{Hales2022}
{Hales}, A.~S., {Marino}, S., {Sheehan}, P.~D., {et~al.} 2022, \apj, 940, 161

\bibitem[{{Hammer} {et~al.}(2017){Hammer}, {Kratter}, \& {Lin}}]{Hammer2017}
{Hammer}, M., {Kratter}, K.~M., \& {Lin}, M.-K. 2017, \mnras, 466, 3533

\bibitem[{{Haugb{\o}lle} {et~al.}(2019){Haugb{\o}lle}, {Weber}, {Wielandt}, {Ben{\'\i}tez-Llambay}, {Bizzarro}, {Gressel}, \& {Pessah}}]{Haugbolle2019}
{Haugb{\o}lle}, T., {Weber}, P., {Wielandt}, D.~P., {et~al.} 2019, \aj, 158, 55

\bibitem[{{Higuchi} {et~al.}(2020){Higuchi}, {K{\'o}sp{\'a}l}, {Mo{\'o}r}, {Nomura}, \& {Yamamoto}}]{Higuchi2020}
{Higuchi}, A.~E., {K{\'o}sp{\'a}l}, {\'A}., {Mo{\'o}r}, A., {Nomura}, H., \& {Yamamoto}, S. 2020, \apj, 905, 122

\bibitem[{{Imaz Blanco} {et~al.}(2023){Imaz Blanco}, {Marino}, {Matr{\`a}}, {Booth}, {Carpenter}, {Faramaz}, {Henning}, {Hughes}, {Kennedy}, {P{\'e}rez}, {Ricci}, \& {Wyatt}}]{ImazBlanco2023}
{Imaz Blanco}, A., {Marino}, S., {Matr{\`a}}, L., {et~al.} 2023, \mnras, 522, 6150

\bibitem[{{Izquierdo} {et~al.}(2021){Izquierdo}, {Testi}, {Facchini}, {Rosotti}, \& {van Dishoeck}}]{Izquierdo2021}
{Izquierdo}, A.~F., {Testi}, L., {Facchini}, S., {Rosotti}, G.~P., \& {van Dishoeck}, E.~F. 2021, \aap, 650, A179

\bibitem[{{Izquierdo} {et~al.}(2023){Izquierdo}, {Testi}, {Facchini}, {Rosotti}, {van Dishoeck}, {W{\"o}lfer}, \& {Paneque-Carre{\~n}o}}]{Izquierdo2023}
{Izquierdo}, A.~F., {Testi}, L., {Facchini}, S., {et~al.} 2023, \aap, 674, A113

\bibitem[{{Kanagawa} {et~al.}(2016){Kanagawa}, {Muto}, {Tanaka}, {Tanigawa}, {Takeuchi}, {Tsukagoshi}, \& {Momose}}]{Kanagawa2016}
{Kanagawa}, K.~D., {Muto}, T., {Tanaka}, H., {et~al.} 2016, \pasj, 68, 43

\bibitem[{{Kanagawa} {et~al.}(2015){Kanagawa}, {Tanaka}, {Muto}, {Tanigawa}, \& {Takeuchi}}]{Kanagawa2015}
{Kanagawa}, K.~D., {Tanaka}, H., {Muto}, T., {Tanigawa}, T., \& {Takeuchi}, T. 2015, \mnras, 448, 994

\bibitem[{{Keppler} {et~al.}(2019){Keppler}, {Teague}, {Bae}, {Benisty}, {Henning}, {van Boekel}, {Chapillon}, {Pinilla}, {Williams}, {Bertrang}, {Facchini}, {Flock}, {Ginski}, {Juhasz}, {Klahr}, {Liu}, {M{\"u}ller}, {P{\'e}rez}, {Pohl}, {Rosotti}, {Samland}, \& {Semenov}}]{Keppler2019}
{Keppler}, M., {Teague}, R., {Bae}, J., {et~al.} 2019, \aap, 625, A118

\bibitem[{{Kley} \& {Dirksen}(2006)}]{Kley2006}
{Kley}, W. \& {Dirksen}, G. 2006, \aap, 447, 369

\bibitem[{{K{\'o}sp{\'a}l} {et~al.}(2013){K{\'o}sp{\'a}l}, {Mo{\'o}r}, {Juh{\'a}sz}, {{\'A}brah{\'a}m}, {Apai}, {Csengeri}, {Grady}, {Henning}, {Hughes}, {Kiss}, {Pascucci}, \& {Schmalzl}}]{Kospal2013}
{K{\'o}sp{\'a}l}, {\'A}., {Mo{\'o}r}, A., {Juh{\'a}sz}, A., {et~al.} 2013, \apj, 776, 77

\bibitem[{{Kral} {et~al.}(2020{\natexlab{a}}){Kral}, {Davoult}, \& {Charnay}}]{Kral2020Nat}
{Kral}, Q., {Davoult}, J., \& {Charnay}, B. 2020{\natexlab{a}}, Nature Astronomy, 4, 769

\bibitem[{{Kral} \& {Latter}(2016)}]{Kral2016b}
{Kral}, Q. \& {Latter}, H. 2016, \mnras, 461, 1614

\bibitem[{{Kral} {et~al.}(2019){Kral}, {Marino}, {Wyatt}, {Kama}, \& {Matr{\`a}}}]{Kral2019}
{Kral}, Q., {Marino}, S., {Wyatt}, M.~C., {Kama}, M., \& {Matr{\`a}}, L. 2019, \mnras, 489, 3670

\bibitem[{{Kral} {et~al.}(2020{\natexlab{b}}){Kral}, {Matr{\`a}}, {Kennedy}, {Marino}, \& {Wyatt}}]{Kral2020}
{Kral}, Q., {Matr{\`a}}, L., {Kennedy}, G.~M., {Marino}, S., \& {Wyatt}, M.~C. 2020{\natexlab{b}}, \mnras, 497, 2811

\bibitem[{{Kral} {et~al.}(2017){Kral}, {Matr{\`a}}, {Wyatt}, \& {Kennedy}}]{Kral2017}
{Kral}, Q., {Matr{\`a}}, L., {Wyatt}, M.~C., \& {Kennedy}, G.~M. 2017, \mnras, 469, 521

\bibitem[{{Kral} {et~al.}(2016){Kral}, {Wyatt}, {Carswell}, {Pringle}, {Matr{\`a}}, \& {Juh{\'a}sz}}]{Kral2016a}
{Kral}, Q., {Wyatt}, M., {Carswell}, R.~F., {et~al.} 2016, \mnras, 461, 845

\bibitem[{{Lagrange} {et~al.}(2019){Lagrange}, {Meunier}, {Rubini}, {Keppler}, {Galland}, {Chapellier}, {Michel}, {Balona}, {Beust}, {Guillot}, {Grandjean}, {Borgniet}, {M{\'e}karnia}, {Wilson}, {Kiefer}, {Bonnefoy}, {Lillo-Box}, {Pantoja}, {Jones}, {Iglesias}, {Rodet}, {Diaz}, {Zapata}, {Abe}, \& {Schmider}}]{Lagrange2019}
{Lagrange}, A.~M., {Meunier}, N., {Rubini}, P., {et~al.} 2019, Nature Astronomy, 3, 1135

\bibitem[{{Law} {et~al.}(2021){Law}, {Loomis}, {Teague}, {{\"O}berg}, {Czekala}, {Andrews}, {Huang}, {Aikawa}, {Alarc{\'o}n}, {Bae}, {Bergin}, {Bergner}, {Boehler}, {Booth}, {Bosman}, {Calahan}, {Cataldi}, {Cleeves}, {Furuya}, {Guzm{\'a}n}, {Ilee}, {Le Gal}, {Liu}, {Long}, {M{\'e}nard}, {Nomura}, {Qi}, {Schwarz}, {Sierra}, {Tsukagoshi}, {Yamato}, {van't Hoff}, {Walsh}, {Wilner}, \& {Zhang}}]{Law2021}
{Law}, C.~J., {Loomis}, R.~A., {Teague}, R., {et~al.} 2021, \apjs, 257, 3

\bibitem[{{Les} \& {Lin}(2015)}]{Les2015}
{Les}, R. \& {Lin}, M.-K. 2015, \mnras, 450, 1503

\bibitem[{{Lieman-Sifry} {et~al.}(2016){Lieman-Sifry}, {Hughes}, {Carpenter}, {Gorti}, {Hales}, \& {Flaherty}}]{Lieman-Sifry2016}
{Lieman-Sifry}, J., {Hughes}, A.~M., {Carpenter}, J.~M., {et~al.} 2016, \apj, 828, 25

\bibitem[{{Lovell} {et~al.}(2021){Lovell}, {Kennedy}, {Marino}, {Wyatt}, {Ansdell}, {Kama}, {Manara}, {Matr{\`a}}, {Rosotti}, {Tazzari}, {Testi}, \& {Williams}}]{Lovell2021}
{Lovell}, J.~B., {Kennedy}, G.~M., {Marino}, S., {et~al.} 2021, \mnras, 502, L66

\bibitem[{{Marino} {et~al.}(2018){Marino}, {Carpenter}, {Wyatt}, {Booth}, {Casassus}, {Faramaz}, {Guzman}, {Hughes}, {Isella}, {Kennedy}, {Matr{\`a}}, {Ricci}, \& {Corder}}]{Marino2018}
{Marino}, S., {Carpenter}, J., {Wyatt}, M.~C., {et~al.} 2018, \mnras, 479, 5423

\bibitem[{{Marino} {et~al.}(2020){Marino}, {Flock}, {Henning}, {Kral}, {Matr{\`a}}, \& {Wyatt}}]{Marino2020}
{Marino}, S., {Flock}, M., {Henning}, T., {et~al.} 2020, \mnras, 492, 4409

\bibitem[{{Marino} {et~al.}(2016){Marino}, {Matr{\`a}}, {Stark}, {Wyatt}, {Casassus}, {Kennedy}, {Rodriguez}, {Zuckerman}, {Perez}, {Dent}, {Kuchner}, {Hughes}, {Schneider}, {Steele}, {Roberge}, {Donaldson}, \& {Nesvold}}]{Marino2016}
{Marino}, S., {Matr{\`a}}, L., {Stark}, C., {et~al.} 2016, \mnras, 460, 2933

\bibitem[{{Marshall} {et~al.}(2023){Marshall}, {Milli}, {Choquet}, {del Burgo}, {Kennedy}, {Kemper}, {Wyatt}, {Kral}, \& {Soummer}}]{Marshall2023}
{Marshall}, J.~P., {Milli}, J., {Choquet}, E., {et~al.} 2023, \mnras, 521, 5940

\bibitem[{{Matr{\`a}} {et~al.}(2017{\natexlab{a}}){Matr{\`a}}, {Dent}, {Wyatt}, {Kral}, {Wilner}, {Pani{\'c}}, {Hughes}, {de Gregorio-Monsalvo}, {Hales}, {Augereau}, {Greaves}, \& {Roberge}}]{Matra2017BetaPic}
{Matr{\`a}}, L., {Dent}, W.~R.~F., {Wyatt}, M.~C., {et~al.} 2017{\natexlab{a}}, \mnras, 464, 1415

\bibitem[{{Matr{\`a}} {et~al.}(2017{\natexlab{b}}){Matr{\`a}}, {MacGregor}, {Kalas}, {Wyatt}, {Kennedy}, {Wilner}, {Duchene}, {Hughes}, {Pan}, {Shannon}, {Clampin}, {Fitzgerald}, {Graham}, {Holland}, {Pani{\'c}}, \& {Su}}]{Matra2017Fomalhaut}
{Matr{\`a}}, L., {MacGregor}, M.~A., {Kalas}, P., {et~al.} 2017{\natexlab{b}}, \apj, 842, 9

\bibitem[{{Matr{\`a}} {et~al.}(2019){Matr{\`a}}, {{\"O}berg}, {Wilner}, {Olofsson}, \& {Bayo}}]{Matra2019}
{Matr{\`a}}, L., {{\"O}berg}, K.~I., {Wilner}, D.~J., {Olofsson}, J., \& {Bayo}, A. 2019, \aj, 157, 117

\bibitem[{{Matr{\`a}} {et~al.}(2015){Matr{\`a}}, {Pani{\'c}}, {Wyatt}, \& {Dent}}]{Matra2015}
{Matr{\`a}}, L., {Pani{\'c}}, O., {Wyatt}, M.~C., \& {Dent}, W.~R.~F. 2015, \mnras, 447, 3936

\bibitem[{{Mo{\'o}r} {et~al.}(2017){Mo{\'o}r}, {Cur{\'e}}, {K{\'o}sp{\'a}l}, {{\'A}brah{\'a}m}, {Csengeri}, {Eiroa}, {Gunawan}, {Henning}, {Hughes}, {Juh{\'a}sz}, {Pawellek}, \& {Wyatt}}]{Moor2017}
{Mo{\'o}r}, A., {Cur{\'e}}, M., {K{\'o}sp{\'a}l}, {\'A}., {et~al.} 2017, \apj, 849, 123

\bibitem[{{Mo{\'o}r} {et~al.}(2019){Mo{\'o}r}, {Kral}, {{\'A}brah{\'a}m}, {K{\'o}sp{\'a}l}, {Dutrey}, {Di Folco}, {Hughes}, {Juh{\'a}sz}, {Pascucci}, \& {Pawellek}}]{Moor2019}
{Mo{\'o}r}, A., {Kral}, Q., {{\'A}brah{\'a}m}, P., {et~al.} 2019, \apj, 884, 108

\bibitem[{{Morbidelli}(2020)}]{Morbidelli2020}
{Morbidelli}, A. 2020, \aap, 638, A1

\bibitem[{{Papaloizou} {et~al.}(2001){Papaloizou}, {Nelson}, \& {Masset}}]{Papaloizou2001}
{Papaloizou}, J.~C.~B., {Nelson}, R.~P., \& {Masset}, F. 2001, \aap, 366, 263

\bibitem[{{Pascucci} {et~al.}(2023){Pascucci}, {Cabrit}, {Edwards}, {Gorti}, {Gressel}, \& {Suzuki}}]{Pascucci2023}
{Pascucci}, I., {Cabrit}, S., {Edwards}, S., {et~al.} 2023, in Astronomical Society of the Pacific Conference Series, Vol. 534, Protostars and Planets VII, ed. S.~{Inutsuka}, Y.~{Aikawa}, T.~{Muto}, K.~{Tomida}, \& M.~{Tamura}, 567

\bibitem[{{Pearce} {et~al.}(2022){Pearce}, {Launhardt}, {Ostermann}, {Kennedy}, {Gennaro}, {Booth}, {Krivov}, {Cugno}, {Henning}, {Quirrenbach}, {Barcucci}, {Matthews}, {Ruh}, \& {Stone}}]{Pearce2022}
{Pearce}, T.~D., {Launhardt}, R., {Ostermann}, R., {et~al.} 2022, \aap, 659, A135

\bibitem[{{Pearce} \& {Wyatt}(2015)}]{Pearce2015}
{Pearce}, T.~D. \& {Wyatt}, M.~C. 2015, \mnras, 453, 3329

\bibitem[{{P{\'e}ricaud} {et~al.}(2017){P{\'e}ricaud}, {Di Folco}, {Dutrey}, {Guilloteau}, \& {Pi{\'e}tu}}]{Pericaud2017}
{P{\'e}ricaud}, J., {Di Folco}, E., {Dutrey}, A., {Guilloteau}, S., \& {Pi{\'e}tu}, V. 2017, \aap, 600, A62

\bibitem[{{Pichierri} {et~al.}(2023){Pichierri}, {Bitsch}, \& {Lega}}]{Pichierri2023}
{Pichierri}, G., {Bitsch}, B., \& {Lega}, E. 2023, \aap, 670, A148

\bibitem[{{Pinilla} {et~al.}(2012){Pinilla}, {Benisty}, \& {Birnstiel}}]{Pinilla2012}
{Pinilla}, P., {Benisty}, M., \& {Birnstiel}, T. 2012, \aap, 545, A81

\bibitem[{{Pinte} {et~al.}(2018{\natexlab{a}}){Pinte}, {M{\'e}nard}, {Duch{\^e}ne}, {Hill}, {Dent}, {Woitke}, {Maret}, {van der Plas}, {Hales}, {Kamp}, {Thi}, {de Gregorio-Monsalvo}, {Rab}, {Quanz}, {Avenhaus}, {Carmona}, \& {Casassus}}]{Pinte2018a}
{Pinte}, C., {M{\'e}nard}, F., {Duch{\^e}ne}, G., {et~al.} 2018{\natexlab{a}}, \aap, 609, A47

\bibitem[{{Pinte} {et~al.}(2018{\natexlab{b}}){Pinte}, {Price}, {M{\'e}nard}, {Duch{\^e}ne}, {Dent}, {Hill}, {de Gregorio-Monsalvo}, {Hales}, \& {Mentiplay}}]{Pinte2018b}
{Pinte}, C., {Price}, D.~J., {M{\'e}nard}, F., {et~al.} 2018{\natexlab{b}}, \apjl, 860, L13

\bibitem[{{Pinte} {et~al.}(2023){Pinte}, {Teague}, {Flaherty}, {Hall}, {Facchini}, \& {Casassus}}]{Pinte2023}
{Pinte}, C., {Teague}, R., {Flaherty}, K., {et~al.} 2023, in Astronomical Society of the Pacific Conference Series, Vol. 534, Protostars and Planets VII, ed. S.~{Inutsuka}, Y.~{Aikawa}, T.~{Muto}, K.~{Tomida}, \& M.~{Tamura}, 645

\bibitem[{{Rebollido} {et~al.}(2022){Rebollido}, {Ribas}, {de Gregorio-Monsalvo}, {Villaver}, {Montesinos}, {Chen}, {Canovas}, {Henning}, {Mo{\'o}r}, {Perrin}, {Rivi{\`e}re-Marichalar}, \& {Eiroa}}]{Rebollido2022}
{Rebollido}, I., {Ribas}, {\'A}., {de Gregorio-Monsalvo}, I., {et~al.} 2022, \mnras, 509, 693

\bibitem[{{Ricci} {et~al.}(2015){Ricci}, {Carpenter}, {Fu}, {Hughes}, {Corder}, \& {Isella}}]{Ricci2015}
{Ricci}, L., {Carpenter}, J.~M., {Fu}, B., {et~al.} 2015, \apj, 798, 124

\bibitem[{{Riviere-Marichalar} {et~al.}(2012){Riviere-Marichalar}, {Barrado}, {Augereau}, {Thi}, {Roberge}, {Eiroa}, {Montesinos}, {Meeus}, {Howard}, {Sandell}, {Duch{\^e}ne}, {Dent}, {Lebreton}, {Mendigut{\'\i}a}, {Hu{\'e}lamo}, {M{\'e}nard}, \& {Pinte}}]{Riviere-Marichalar2012}
{Riviere-Marichalar}, P., {Barrado}, D., {Augereau}, J.~C., {et~al.} 2012, \aap, 546, L8

\bibitem[{{Roberge} {et~al.}(2013){Roberge}, {Kamp}, {Montesinos}, {Dent}, {Meeus}, {Donaldson}, {Olofsson}, {Mo{\'o}r}, {Augereau}, {Howard}, {Eiroa}, {Thi}, {Ardila}, {Sandell}, \& {Woitke}}]{Roberge2013}
{Roberge}, A., {Kamp}, I., {Montesinos}, B., {et~al.} 2013, \apj, 771, 69

\bibitem[{{Rosenfeld} {et~al.}(2013){Rosenfeld}, {Andrews}, {Hughes}, {Wilner}, \& {Qi}}]{Rosenfled2013}
{Rosenfeld}, K.~A., {Andrews}, S.~M., {Hughes}, A.~M., {Wilner}, D.~J., \& {Qi}, C. 2013, \apj, 774, 16

\bibitem[{{Schneiderman} {et~al.}(2021){Schneiderman}, {Matr{\`a}}, {Jackson}, {Kennedy}, {Kral}, {Marino}, {{\"O}berg}, {Su}, {Wilner}, \& {Wyatt}}]{Schneiderman2021}
{Schneiderman}, T., {Matr{\`a}}, L., {Jackson}, A.~P., {et~al.} 2021, \nat, 598, 425

\bibitem[{Shakura \& Sunyaev(1973)}]{Shakura1973}
Shakura, N.~I. \& Sunyaev, R.~A. 1973, in IAU Symposium, Vol.~55, X- and Gamma-Ray Astronomy, ed. H.~{Bradt} \& R.~{Giacconi}, 155

\bibitem[{{Su} {et~al.}(2013){Su}, {Rieke}, {Malhotra}, {Stapelfeldt}, {Hughes}, {Bonsor}, {Wilner}, {Balog}, {Watson}, {Werner}, \& {Misselt}}]{Su2013}
{Su}, K. Y.~L., {Rieke}, G.~H., {Malhotra}, R., {et~al.} 2013, \apj, 763, 118

\bibitem[{{Teague} {et~al.}(2018){Teague}, {Bae}, {Bergin}, {Birnstiel}, \& {Foreman-Mackey}}]{Teague2018}
{Teague}, R., {Bae}, J., {Bergin}, E.~A., {Birnstiel}, T., \& {Foreman-Mackey}, D. 2018, \apjl, 860, L12

\bibitem[{{Visser} {et~al.}(2009){Visser}, {van Dishoeck}, \& {Black}}]{Visser2009}
{Visser}, R., {van Dishoeck}, E.~F., \& {Black}, J.~H. 2009, \aap, 503, 323

\bibitem[{{Wyatt}(2008)}]{Wyatt2008}
{Wyatt}, M.~C. 2008, \araa, 46, 339

\bibitem[{{Zhang} {et~al.}(2021){Zhang}, {Booth}, {Law}, {Bosman}, {Schwarz}, {Bergin}, {{\"O}berg}, {Andrews}, {Guzm{\'a}n}, {Walsh}, {Qi}, {van't Hoff}, {Long}, {Wilner}, {Huang}, {Czekala}, {Ilee}, {Cataldi}, {Bergner}, {Aikawa}, {Teague}, {Bae}, {Loomis}, {Calahan}, {Alarc{\'o}n}, {M{\'e}nard}, {Le Gal}, {Sierra}, {Yamato}, {Nomura}, {Tsukagoshi}, {P{\'e}rez}, {Trapman}, {Liu}, \& {Furuya}}]{Zhang2021}
{Zhang}, K., {Booth}, A.~S., {Law}, C.~J., {et~al.} 2021, \apjs, 257, 5

\bibitem[{{Zhang} {et~al.}(2018){Zhang}, {Zhu}, {Huang}, {Guzm{\'a}n}, {Andrews}, {Birnstiel}, {Dullemond}, {Carpenter}, {Isella}, {P{\'e}rez}, {Benisty}, {Wilner}, {Baruteau}, {Bai}, \& {Ricci}}]{Zhang2018}
{Zhang}, S., {Zhu}, Z., {Huang}, J., {et~al.} 2018, \apjl, 869, L47

\bibitem[{{Zhu}(2022)}]{Zhu2022}
{Zhu}, W. 2022, \aj, 164, 5

\bibitem[{{Zhu} \& {Zhang}(2022)}]{ZhuZhang2022}
{Zhu}, Z. \& {Zhang}, R.~M. 2022, \mnras, 510, 3986

\end{thebibliography}

\begin{appendix}

\begin{landscape}

\section{Characteristics of the known gas-bearing debris disks}
\label{appendix_knowndisks}

\begin{table}[h]            
\centering                          
\caption{Characteristics of the known gas-bearing debris disks. } 
\begin{tabular}{c c c c c c c c c c c c}        
\hline             
 & & & & & & & Integrated flux & & & & \\
Target & Spectral & Distance & $\rm T_{eff}$ & $\rm M_*$ & Age & Inclination & CO flux & Ref. & $\rm M_{CO}$ & Ref. & C, O ?  \\
 & type & [pc] & [K] & [$\rm M_\odot$] & [Myr] & [$^\circ$] & [Jy.km/s] & & [$\rm M_\oplus$] & & \\ 
\hline                 
    $\beta$ Pic  & A6  & 19.44 $\pm$ 0.05     & 8200  & $1.73^{+0.00}_{-0.02}$ & 23-29                & 88         & 4.5     & (1)  & $3.60\times 10^{-5}$ & (1) & CI(2), CII(3), OI(4) \\
    49 Ceti      & A1  & 57.233 $\pm$ 0.1792  & 8900  & $2.02^{+0.03}_{-0.05}$ & 40-50                & 81         & 3.87    & (5)  & $1.46\times 10^{-1}$ & (6) & CI(2), CII(7), OI(7) \\
    Fomalhaut    & A3  & 7.7 $\pm$ 0.03       & 8900  & $1.8^{+0.07}_{-0.06}$  & 564-916              & 66         & 0.068   & (8)  & $5.22\times 10^{-7}$ & (8) & $-$ \\
    TWA 7        & M3  & 34.097 $\pm$ 0.0317  & 3415* & $0.46^{+0.07}_{-0.1}$  & $6.4^{+1.0}_{-1.2}$  & 10         & 0.091** & (9)  & $8.00\times 10^{-6}$ & (9) & $-$ \\  
    NO Lup       & K7  & 132.896 $\pm$ 0.2932 & 3994* & 0.7                    & 1-3                  & $50\pm 30$ & 0.29**  & (10) & $4.9 \times 10^{-5}$ & (10)& $-$ \\ 
    HD 21997     & A3  & 69.686 $\pm$ 0.1374  & 8433* & 1.7                    & 30-45                & 32         & 2.17    & (11) & $2.16\times 10^{-1}$ & (6) & CI(2) \\  
    HD 32297     & A5/7& 129.734 $\pm$ 0.5453 & 7700  & $1.69^{+0.02}_{-0.02}$ & 15-30                & 78         & 1.05    & (5)  & $3.16\times 10^{-1}$ & (12)& CI(2), CII(13) \\      
    HD 36546     & B8  & 100.179 $\pm$ 0.4174 & $-$   & $-$                    & 3-10                 & 78         & 2.67    & (14) & $3.20\times 10^{-3}$ & (14)& $-$ \\ 
    HD 95086     & A8  & 86.461 $\pm$ 0.1398  & 7600  & $1.61^{+0.02}_{-0.01}$ & 12-18                & 30         & 0.0095  & (15) & $4.27\times 10^{-6}$ & (15)& $-$ \\ 
    HD 110058    & A0  & 130.076 $\pm$ 0.5313 & 8000  & $1.7^{+0.03}_{-0.02}$  & 12-18                & 85         & 0.091   & (16) & $6.90\times 10^{-2}$ & (16)& $-$ \\
    HD 121191    & A5  & 132.286 $\pm$ 0.4497 & 7690* & 1.6                    & 15-16                & 28         & 0.21    & (17) & $2.30\times 10^{-3}$ & (17)& CI(2)\\
    HD 121617    & A1  & 117.890 $\pm$ 0.4489 & 9285* & 1.9-2.27               & 16                   & 37-44      & 1.27    & (18) & $1.80\times 10^{-2}$ & (18)& CI(2)\\
    HD 129590    & G1  & 136.322 $\pm$ 0.4404 & 5910  & $1.4^{+0.02}_{-0.01}$  & 14-18                & 65-70      & 0.056   & (17) & $4.47\times 10^{-5}$ & (17)& $-$ \\
    HD 131488    & A1  & 152.244 $\pm$ 0.8460 & 9000* & 1.88                   & 16                   & 82         & 0.78    & (18) & $8.90\times 10^{-2}$ & (18)& $-$ \\
    HD 131835    & A2  & 129.739 $\pm$ 0.4679 & 8100  & $1.77^{+0.05}_{-0.04}$ & 14-18                & 79         & 0.798   & (19) & $4.00\times 10^{-2}$ & (20)& CI(2)\\
    HD 138813    & A0  & 136.597 $\pm$ 0.5859 & 8100  & $2.15^{+0.07}_{-0.09}$ & 7-13                 & 29         & 1.406   & (21) & $3.10\times 10^{-3}$ & (19)& $-$ \\
    HD 141569    & A2  & 111.611 $\pm$ 0.3650 & 8400  & $2.04^{+0.04}_{-0.07}$ & 2-8                  & 57         & 10      & (22) & $1.10\times 10^{-1}$ & (22)& $-$ \\
    HD 146897    & F2  & 132.186 $\pm$ 0.4141 & 6200  & $1.28^{+0.02}_{-0.01}$ & 7-13                 & 89         & 0.06    & (21) & $1.00\times 10^{-2}$ & (21)& $-$ \\
    HD 156623    & A0  & 108.334 $\pm$ 0.3286 & 8350  & $1.9^{+0.04}_{-0.01}$  & 14-18                & 31.8       & 1.183   & (21) & $2.45\times 10^{-3}$ & (19)& $-$ \\
    HD 172555    & A7  & 28.789 $\pm$ 0.1305  & 7846* & 1.76                   & 23                   & 102        & 0.12    & (23) & $7.50\times 10^{-6}$ & (23)& OI(24) \\
    HD 181327    & F5/6& 47.777 $\pm$ 0.0653  & 6400  & $1.39^{+0.01}_{-0.01}$ & 23-29                & 30         & 0.0301  & (25) & $1.87\times 10^{-6}$ & (25)& $-$  \\    

\hline
    
\end{tabular}

{\raggedright \vspace{3mm} \textbf{Notes}: All distances originate from the \citealp{GaiaDR3}. All the stellar properties have been gathered from \citealp{Esposito2020}, except the one marked by a star (*) where the properties are gathered from the \href{http://simbad.u-strasbg.fr/simbad/}{\texttt{SIMBAD}} catalog with the most recent values to date. 
The references for the fluxes listed in the integrated flux column are listed in the next column. The fluxes are originating from the CO(J = 2-1) (Band 6) observations, except for the ones marked with a double star (**), that were only observed in Band 7, probing the CO(J = 3-2) emission line. The references for each CO masses is listed in the next column. The references for each detection of other molecules are noted in parenthesis. \\ 

\vspace{2mm}

\textbf{References}: (1) \citealp{Matra2017BetaPic}; (2) \citealp{Cataldi2023}; (3) \citealp{Cataldi2014}; (4) \citealp{Brandeker2016}; (5) \citealp{Moor2019}; (6) \citealp{Higuchi2020}; (7) \citealp{Roberge2013}; (8) \citealp{Matra2017Fomalhaut}; (9) \citealp{Matra2019}; (10) \citealp{Lovell2021}; (11) \citealp{Kospal2013}; (12) \citealp{Cataldi2020}; (13) \citealp{Donaldson2013}; (14) \citealp{Rebollido2022}; (15) \citealp{Booth2019}; (16) \citealp{Hales2022}; (17) \citealp{Kral2020}; (18) \citealp{Moor2017}; (19) \citealp{Hales2019}; (20) \citealp{Kral2019}; (21) \citealp{Lieman-Sifry2016}; (22) \citealp{DiFolco2020}; (23) \citealp{Schneiderman2021}; (24) \citealp{Riviere-Marichalar2012}; (25) \citealp{Marino2016} \par}

\label{tab:known_disks}
\end{table}

\end{landscape}

\clearpage

\section{Hydrodynamic simulations}
\label{appendix_hydro}

\begin{figure*}[t]
   \centering   
   \includegraphics[scale=0.33]{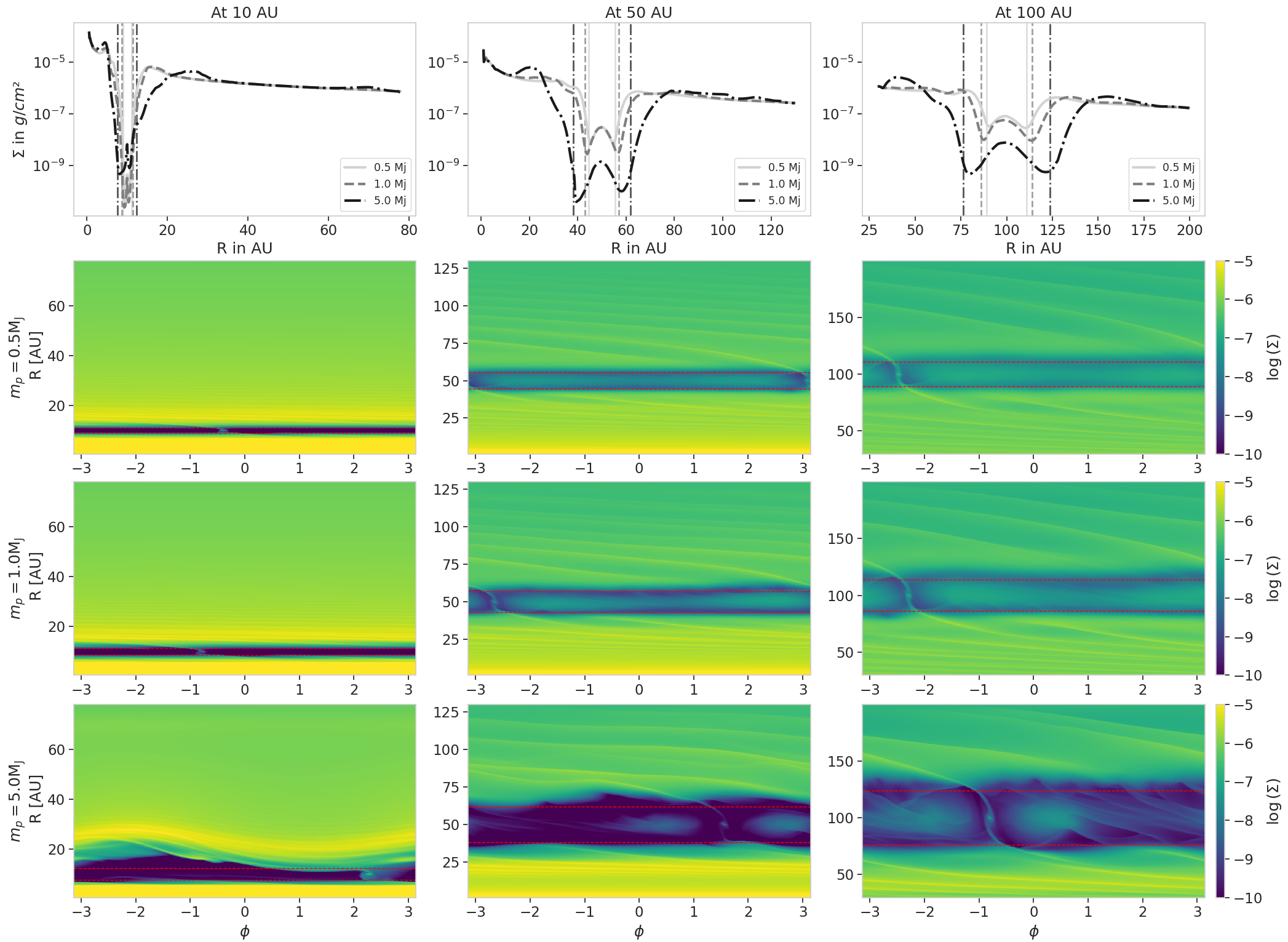}
   \caption{Surface density distributions of the gas in our hydrodynamical simulations. In the top row, we show the azimuthally averaged profiles for the  planets located at 10AU (first column), 50AU (second column) and 100AU (third column). The color and style of the lines show the different planet masses. The vertical matching lines are located at $r_p \pm 2r_H$ in order to estimate the gap width. We clearly see that most massive planets produce deeper and wider gaps. The remaining panels show the 2D density maps for the different planet semi-major axes (columns) and masses (rows). The dashed red lines show again the gap width based on the planet's Hill sphere. The $5 \; \rm M_J$ planet located at 10AU produces an eccentric gap, which produces a less deep gap than in the circular cases. }
   \label{gap_widths_fargo}
\end{figure*}

We present in this appendix the outputs of our hydrodynamical simulations used to derive the synthetic images shown in this study. In Fig. \ref{gap_widths_fargo} we show in the first row the azimuthally averaged radial profiles of the disks at t = $10^4$ orbits. The vertical lines show the gap width as estimated from the Hill sphere of the planet ($w_{gap} = 4 r_H$, centered on $r_p$) in each configuration (see Sect. \ref{subsec:CASA}). The profiles for the three different planet masses are over-plotted in the same graph and the different planet distances are shown in the different columns. Below, we show the corresponding 2D surface density maps for each planet mass (increasing from top to bottom). The red dashed lines are located at $r_p \pm 2 r_H$ to highlight the different gap widths.

Here, we can see that the planets present different gap depths, with more or less corotating material (see Sect. \ref{sec:discuss_sigmagap}). It is also very clear that the $5 \; \rm M_J$ planet located at 10AU produces an eccentric gap \citep[e.g.,][]{Kley2006,Bitsch2013eccentricity}. We can also distinguish the spiral arms responsible for the velocity deviations producing kinks. 

The estimate of the gap width from the Hill sphere of the planet matches quite well the gap width resulting from the hydrodynamical simulations. However, one can note that this estimate matches the width at the bottom of the gap and the steepness of the gap edges depend on both the gas viscosity and planet mass. Therefore, in order to properly use Eq. \ref{eq_gap_width}, one needs to use the width at the bottom of the gap. 

In \cite{Kanagawa2016}, the authors derive a way to link the gap width to the planet mass. However, we find that the gap width resulting from our hydrodynamical simulations is better represented by the Hill sphere estimate than their estimate (see also the discussion about the gap depth in Sect. \ref{sec:discuss_sigmagap}). In order to be coherent with our simulations, we therefore use the Hill sphere.

\section{Radial profiles in all configurations}
\label{appendix_profiles}

\begin{figure*}[t]
   \centering   
   \includegraphics[scale=0.29]{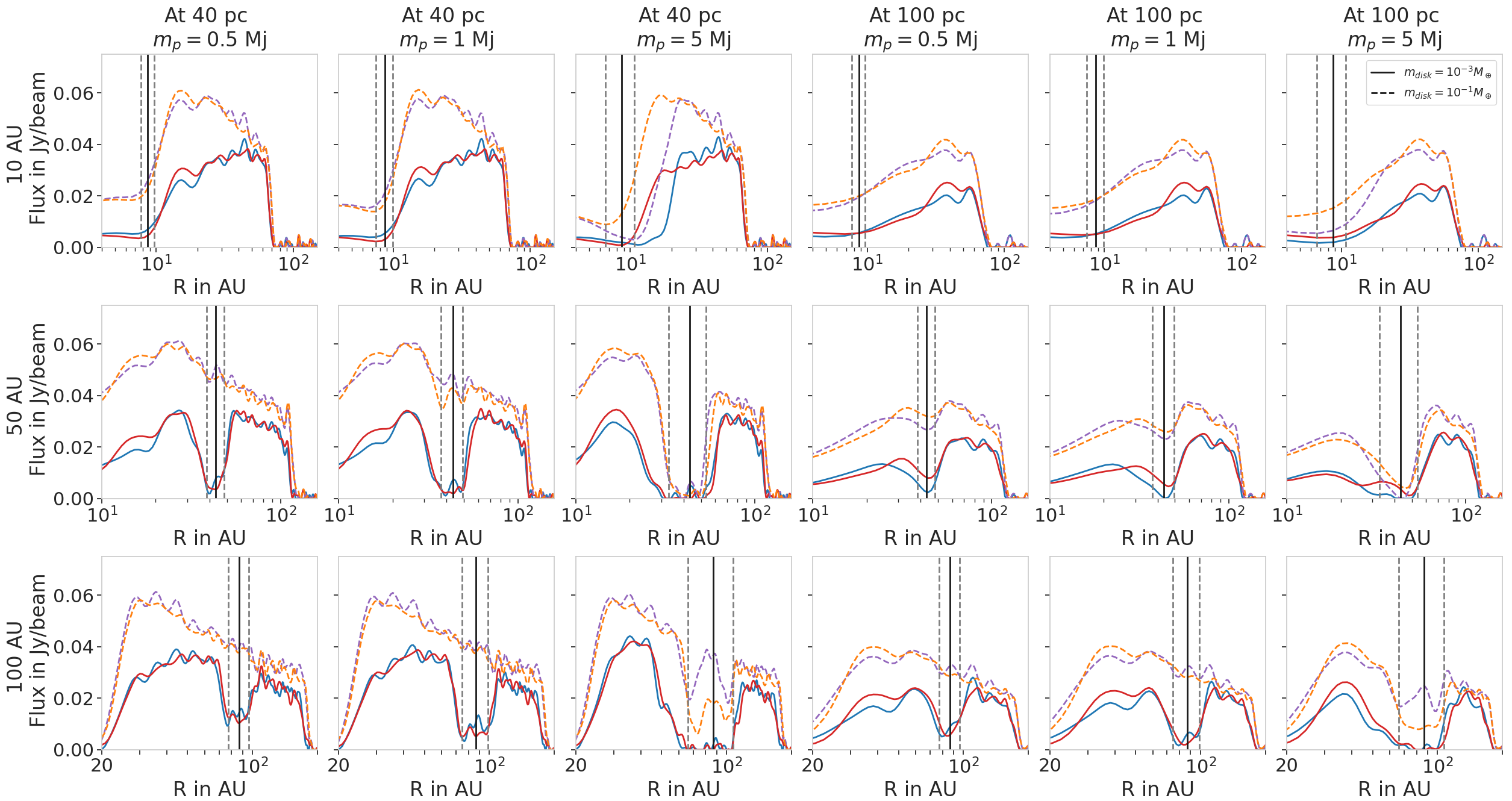}
   \caption{Radial profiles along two axes as in Fig. \ref{profile_1em3} for all the planet configurations. The planet mass increases from left to right and the planet location increases from top to bottom. The first (resp. last) three columns show the profiles taken from the images of the disks located at 40 pc (resp. 100 pc). Both disk masses are over-plotted: the red and blue lines represent the low mass disks whereas the dashed orange and purple lines represent the high mass disks. The vertical black line shows the planet location (projected due to the disk's inclination) and the vertical dashed gray lines show the gap width. Here, we clearly see that some gaps seen in the low mass disks are not visible in heavier disks.}
   \label{profiles_all_config}
\end{figure*}

In Sect. \ref{sec:diffplanetmass} we show that planetary gaps are observable in the gas emission of debris disks with $m_{disk} = 10^{-3} M_\oplus$. Radial profiles allow us to look at these gaps from another point of view. In Fig. \ref{profiles_all_config}, we show these radial profiles in all the planetary configurations for two disk masses and disk distances. 

We clearly see that the gaps are visible when the total disk mass is lower. The gap width (represented by the vertical dashed gray lines) matches the size of the gap seen in the images. A few exceptions arise when the planet is heavy and therefore becomes eccentric: an eccentric planet will produce a wider gap with smoother gap edges compared to a circular planet. 

In the 10 AU case, the gaps are too close to the central star to be distinguished with the investigated resolution (see Sect. \ref{subsec:CASA}). Moreover, when the disk is located further away (three last columns of Fig. \ref{profiles_all_config}), the gaps are more difficult to resolve. However, we clearly see that here, when the disks are located at 100 pc, gaps located at 50AU and 100AU are still observable, even when they are produced by low mass giant planets ($m_p = 0.5 \; \rm M_J$).

\section{Images at different disk distances}
\label{appendix_distances}

\begin{figure*}[t]
   \centering   
   \includegraphics[scale=0.75]{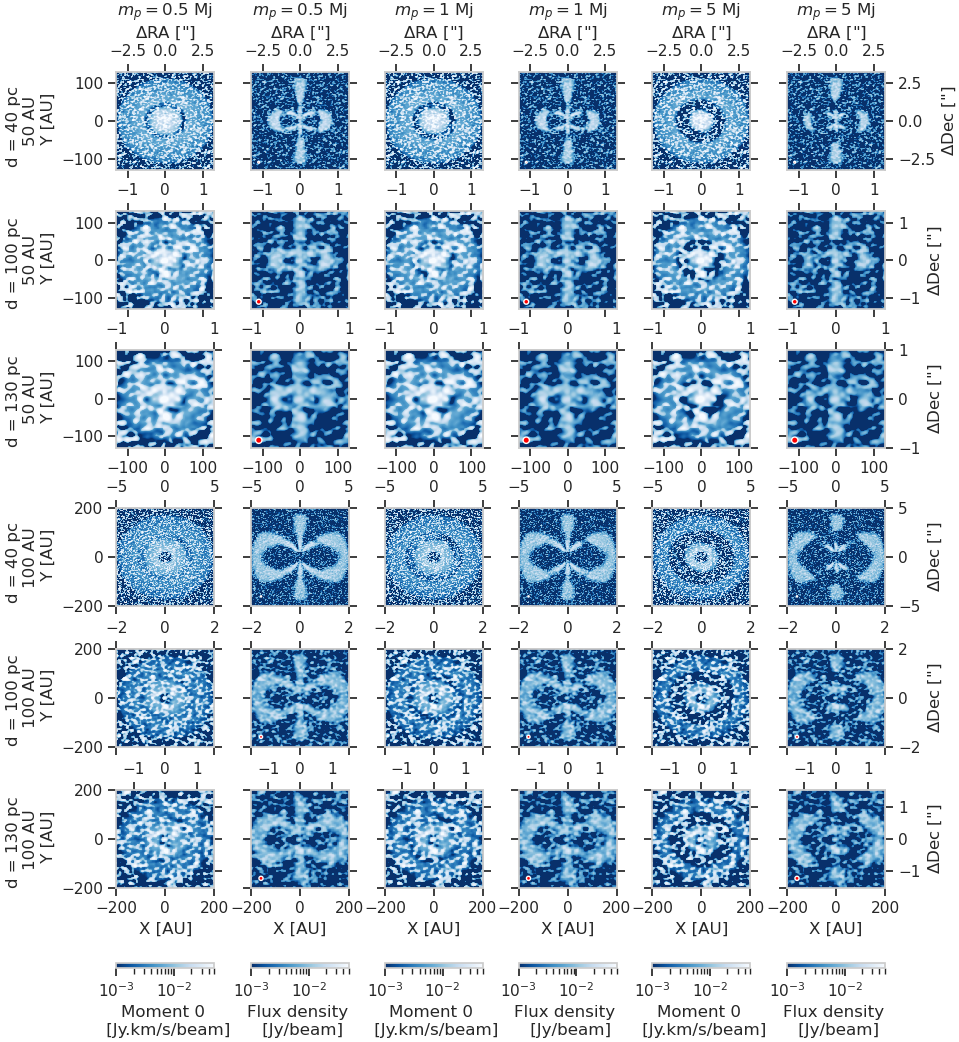}
   \caption{Moment 0 (1st, 3rd and 5th columns) and combined channel maps at dv = -1.0; 0; 1.0 km/s (2nd, 4th and 6th columns) for different configurations: the planet mass increases from left to right; the 3 first rows correspond to the planet located at 50 AU, while in the 3 last one the planet is located at 100 AU; for each planet location, the disk distance is increased from top to bottom (40pc, 100pc, 130pc). We show the channel maps issued from \texttt{CASA} in Jy/beam with the resulting beam shown in red in the bottom left corner. Here, the disk's mass is $\rm 10^{-3} \; M_\oplus$ of CO. The observability of the planetary gaps depends on whether the gap is resolved or not.}
   \label{images_diffdiskdist}
\end{figure*}

We show in Fig. \ref{images_diffdiskdist} the synthetic images of our simulated disks when they are located at 40pc, 100pc and 130pc. As mentioned in Sect. \ref{subsubsec:discuss_diskdist}, the distance of the disk will impact both our ability to resolve the gap and to detect the gas outside of the gap. 

\end{appendix}


\end{document}